\documentclass[a4paper,fleqn,usenatbib]{mnras}

\usepackage{newtxtext,newtxmath}

\usepackage[T1]{fontenc}
\usepackage{ae,aecompl}

\usepackage{graphicx}	
\usepackage{amsmath}	
\usepackage[export]{adjustbox}

\title[Radio detections of runaway stellar bow shocks]{Radio detections of IR-selected runaway stellar bow shocks}

\author[Van den Eijnden et al.]
{J. van den Eijnden$^{1}$\thanks{E-mail: jakob.vandeneijnden@st-hildas.ox.ac.uk}, P. Saikia$^{2}$ and S. Mohamed$^{3,4,5}$
\\
$^{1}$Astrophysics, Department of Physics, University of Oxford, Denys Wilkinson Building, Keble Road, Oxford OX1 3RH, UK\\
$^{2}$Center for Astro, Particle and Planetary Physics (CAP$^3$), New York University Abu Dhabi, PO Box 129188, Abu Dhabi, UAE\\
$^{3}$South African Radio Astronomy Observatory, 2 Fir Street, Observatory, 7925, South Africa\\
$^{4}$Department of Astronomy, University of Cape Town, Private Bag X3, Rondebosch 7701, South Africa\\
$^{5}$National Institute for Theoretical and Computational Sciences (NITheCS), KwaZulu-Natal, South Africa\\
}

\date{Accepted XXX. Received YYY; in original form ZZZ}

\pubyear{2021}

\begin{document}
\label{firstpage}
\pagerange{\pageref{firstpage}--\pageref{lastpage}}
\maketitle

\begin{abstract}
Massive stars moving at supersonic peculiar velocities through the interstellar medium (ISM) can create bow shocks, arc-like structures at the interface between the stellar wind and the ISM. Many such bow shocks have been detected and catalogued at IR wavelengths, but detections in other wavebands remain rare. Strikingly, while electrons are expected to be accelerated in the bow shock and their non-thermal emission may include synchrotron emission at low frequencies, only two massive runaway stellar bow shocks have to date been detected in the radio band. Here, we examine a sample of fifty IR-detected bow shocks from the E-BOSS catalogues in recently released radio images from the Rapid ASKAP Continuum Survey (RACS). We identify three confident and three likely counterparts, as well as three inconclusive candidates requiring confirmation via follow-up observations. These detections significantly increase the number of known radio massive stellar bow shocks and highlight the advantage of dedicated searches with current and next-generation radio telescopes. We investigate the underlying radio emission mechanism for these radio sources, finding a mix of free-free-dominated and synchrotron-dominated systems. We also discuss the non-detected targets by putting constraints on their emission properties and investigating their detectability with future observations. Finally, we propose several future avenues of research to advance the study and understanding of bow shocks at radio frequencies.
\end{abstract}

\begin{keywords}
stars: early-type -- stars: individual: -- radio continuum: general -- shock waves
\end{keywords}



\section{Introduction}

Runaway massive stars are usually early-type stars showing large peculiar velocities, likely caused either by a kick from the supernova explosion of a binary companion \citep{zwicky1957,blaauw1961} or by dynamical ejection after gravitational interactions with other stars \citep{poveda1967,gies1986}. If the runaway star's peculiar velocity significantly exceeds the local interstellar medium (ISM) sound speed, it can create a bow shock in the direction of its motion. As these early-type stars tend to launch powerful stellar winds, the bow shock is located at the distance from the star where the wind and ISM ram pressure are in equilibrium \citep{baranov1971}. The typical arc-type shape of the bow shock is set by the combination of stellar (wind) properties -- i.e. wind and peculiar velocity, mass-loss rate -- as well as the local ISM density, the presence of complex structures, and viewing angle \citep{wilkin1996}. 

The bow shocks of runaway massive stars can, in theory, emit brightly across the entire spectrum via various emission mechanisms. The largest number of bow shocks has been identified at IR wavelengths, where the emission of shock-heated dust peaks \citep[see e.g.][for catalogues]{peri2012,peri2015,kobulnicky2016}. Line emission, for instance in H$\alpha$, has been detected for a subset of these samples \citep[e.g.][]{gull1979,kaper1997}. Such line emission highlights the presence of a thermal electron population that may similarly contribute to low-frequency (especially radio) emission via thermal free-free processes. Alternatively, at the shock, diffusive shock acceleration of electrons could create an electron population reaching energies up to the TeV regime \citep[e.g.][]{delvalle2012}. The non-thermal emission of this population may contribute both at high and low frequencies, respectively via inverse Compton scattering at X-ray energies and beyond, and via synchrotron emission at radio frequencies. To date, however, at high frequencies, no bow shock emission has been detected unambiguously \citep{schulz2014,toala2016,toala2017,debecker2017,hess2018}, although two unidentified \textit{Fermi} sources may be associated with bow shocks \citep{sanchezayaso2018}. 

Searches of synchrotron radio emission could therefore offer a fruitful alternative to study bow shock particle acceleration. At the time of writing, low-frequency bow shock studies have only been slightly more successful, as only two runaway stellar bow shocks have been confidently detected at radio frequencies. The prototype radio bow shock was discovered by \citet{benaglia2010} around BD+43$^{\rm o}$3654 and further observed by \citet{brookes2016} and \citet{benaglia2021}. More recently, MeerKAT observations of the X-ray binary Vela X-1 discovered radio emission from its known bow shock \citep{vandeneijnden2021}. While the former detection has been identified as synchrotron emission, providing an insight into the non-thermal electron population, the latter source's radio emission appears to be dominated by thermal free-free emission. Therefore, radio studies of more runaway bow shocks are paramount to understand how often and in what circumstances these sources efficiently accelerate electrons to the point where their non-thermal emission becomes detectable and dominates the low-frequency spectrum. While existing VLA and ATCA data has revealed a handful of possible radio sources at bow shock positions in survey data \citep{peri2012,peri2015,benaglia2013}, the sensitivity and spatial resolution of those observations was typically insufficient to confidently assign the radio emission to the bow shock. 

The advent of precursors to the Square Kilometer Array (SKA), such as MeerKAT and the Australian SKA Pathfinder (ASKAP) presents an opportunity to search for stellar bow shock radio counterparts at high sensitivity to extended emission. The recent first data release of the Rapid ASKAP Continuum Survey \citep[RACS;][]{mcconnell2020}, highlights this development: RACS presents radio images of the entire Southern and part of the Northern sky (i.e. declinations below $+41^{\rm o}$, at the time of writing) with typical RMS sensitivities of $\sim 0.25$ mJy/beam. In addition to its sky coverage and sensitivity, its typical resolution of $\sim15$ arcseconds improves upon earlier surveys, facilitating a more effective search for bow shock morphologies. Building on the radio detection of Vela X-1, we therefore searched for radio counterparts of known IR runaway bow shocks in the first and second version of the E-BOSS catalogue \citep{peri2012,peri2015}. 

In this Paper, we report on the radio detection of three radio bow shock counterparts, as well as three strong candidates and three to-be-confirmed counterparts, which all warrant follow up studies. In Section \ref{sec:data}, we will describe the public radio and IR survey data used in our study, while in Section \ref{ref:search}, we will discuss all E-BOSS sources where radio emission is identified in RACS. In Section \ref{sec:calcs}, we perform basic analytical calculations to assess whether the radio emission can be explained in existing bow shock frameworks and to better understand whether the radio sources are dominated by bow shock emission. We discuss our findings and turn to future prospects for radio bow shock studies in Section \ref{sec:5}, before finally concluding in Section \ref{sec:conc}. 

\section{Observations and target selection}
\label{sec:data}

We accessed the publicly-available RACS images of the fields containing each of the investigated IR bow shocks via the CSIRO ASKAP Science Data Archive (CASDA; \url{https://research.csiro.au/racs/home/data/}). The RACS data were taken with the full ASKAP array, consisting of $36$ antennas equipped with a phased array feed to digitally sample a large, 31 square degree field of a view. The images in the first data release are taken at a central frequency of $887.5$ MHz with a bandwidth of $288$ MHz. For details on ASKAP, we refer the reader to \citet{hotan2014}, \citet{hotan2021}, and \citet{mcconnell2016}, while more details about RACS, its first data release, and a comparison with pre-existing radio surveys can be found in \citet{mcconnell2020}. For multi-wavelength comparison of the radio images, we consulted the WISE infrared images in the NASA/IPAC Infrared Science Archive (\url{https://irsa.ipac.caltech.edu/applications/wise/}) and SuperCOSMOS H$\alpha$ images \citep[][\url{http://www-wfau.roe.ac.uk/sss/halpha/}]{parker2005}.

For this first bow shock search in RACS, we selected sources in the first and second E-BOSS catalogues \citep[short for Extensive stellar BOw Shock Survey;][]{peri2012,peri2015}. These catalogues list bow shock and bow shock candidates detected in IR with, predominantly, MSX and WISE. We selected all sources covered by the first data release of RACS, creating a sample of 50 targets. In addition to presenting IR images, the E-BOSS catalogues also measure and list the geometrical properties of the identified bow shocks and use those in combination with stellar (wind) parameters to infer the surrounding ISM density. These estimates are especially useful for our inferences regarding the nature of any detected radio emission (see Section \ref{sec:calcs}). \citet{peri2012} and \citet{peri2015} also note that a number of sources show hints of radio emission in either NVSS \citep{condon1998} or ATCA images; however, the typical $\sim 0.5$ mJy/beam noise level in combination with differences between radio and IR morphology yielded these possible radio counterparts challenging to interpret further at the time. Since several of these targets are covered by RACS, which has a higher sensitivity and spatial resolution, we will discuss them in extra detail in this work. Finally we note that, based on \textit{Spitzer} observations, \citet{kobulnicky2016} present a larger IR catalogue, that can be explored in follow-up work. Here, we focus on the sources in the E-BOSS samples with listed geometrical and ISM properties.

\section{RACS bow shock search}
\label{ref:search}

\begin{table*}
\caption{Summary of our morphological assessment of the 10 radio counterparts of the E-BOSS IR bow shocks covered by RACS. All other bow shocks in E-BOSS have either not been covered by, or do not show radio emission in, RACS. Here, we list the source name and associated star-forming region (if relevant), the assessment, the main argument and complications discussed in the text, and finally the figures related to the source. The sources are discussed in the same order in the Results section. BS is short-hand for bow shock. *\textit{While K5 and HIP 38430 are listed as likely counterparts based on their morphology, we refer the reader to Sections \ref{sec:calcs} and \ref{sec:5} for (possible) issues in explaining their radio brightness through feasible physical models.}}
\label{tab:assessment}
\begin{tabular}{llllll}
\hline
Source & Field & Radio BS? & Reasoning & Complications & Figures \\ \hline
G1 & NGC 6357 & Yes & Morphology, position & Diffuse structures in field; bright emission from NGC 6357 & \ref{fig:Gfield}, \ref{fig:Gfield_radvswise} (top) \\
G3 & NGC 6357 & Yes & Morphology, position & -- & \ref{fig:Gfield}, \ref{fig:Gfield_radvswise} (middle) \\ \hline
S1 & RCW 49 & Yes & Morphology, position & Source extension towards the North & \ref{fig:RCW49_field}, \ref{fig:RCW49field_radvswise} (top) \\
S2 & RCW 49 & Unclear & Position & Part of large-scale emission of RCW 49 & \ref{fig:RCW49_field}, \ref{fig:RCW49field_radvswise} (bottom left) \\
S3 & RCW 49 & Unclear & Position, morphology? & Part of large-scale emission of RCW 49 & \ref{fig:RCW49_field}, \ref{fig:RCW49field_radvswise} (bottom right) \\ \hline
HIP 88652 & -- & TBC & Morphology, position & Large extension; residuals from close-by point sources & \ref{fig:HIP88652_K5_field_radvswise} (top) \\
K5* & Cyg OB2 & Likely & Morphology, position & Diffuse radio source to the South & \ref{fig:HIP88652_K5_field_radvswise} (bottom) \\
HIP 98418 & -- & Likely & Morphology, position & Part of larger diffuse structure & \ref{fig:HIP98418_38430field_radvswise} (top) \\
HIP 38430* & -- & Likely & Morphology, position & Part of larger diffuse structure & \ref{fig:HIP98418_38430field_radvswise} (top) \\
HIP 24575 & -- & No & Point source & No common morphology, likely background source & \ref{fig:HIP24575_field_radvswise} \\ \hline
\end{tabular}\\
\end{table*}

For each E-BOSS IR bow shock covered by a RACS field, we started by visually inspecting the source position in the radio image. For the majority of sources, no point source or extended emission was visible at the source position, yielding instead typical $3-\sigma$ upper limits on the peak flux density of $\sim 0.5$-$1$ mJy. However, for ten IR-detected bow shocks, radio emission can be seen at or close to its position in RACS. In this section, we first walk through these ten targets and then discuss whether this radio emission likely originates from the bow shock, or instead from other, unrelated diffuse structures or background sources. We summarize our assessment and classification of each of the ten considered sources in Table \ref{tab:assessment}. There, we also list their related Figures for clarity. The classifications in Table \ref{tab:assessment} are purely based on morphology -- we will include physical calculations in our assessment in later sections. Before discussing individual sources, we stress that any region drawn in the figures in this paper is intended to highlight the discussed features and to compare radio and IR images -- they are not used in the actual calculations.

\begin{figure}
 \includegraphics[width=\columnwidth]{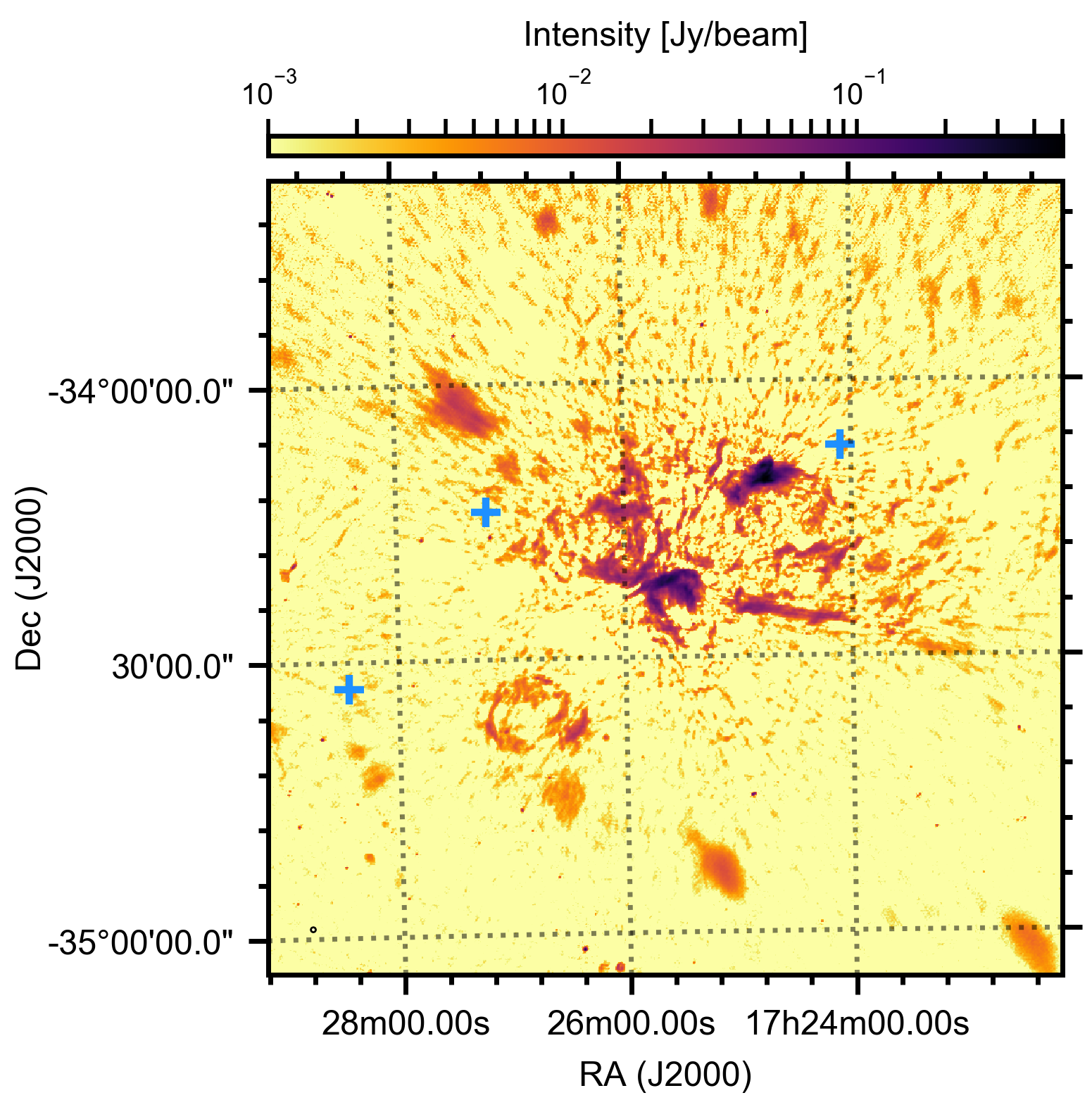}
 \caption{Logaritmically-scaled RACS image of the star-forming region NGC 6357. The position of three known IR bow shocks, G1, G3, and G8, are indicated by the crosses. Zooms of these positions are shown in Figure \ref{fig:Gfield_radvswise}. The complex field also reveals how the RMS sensitivity can vary significantly with position, explaining why we determine local upper limits for radio non-detected bow shocks.}
 \label{fig:Gfield}
\end{figure}

\begin{figure*}
 \includegraphics[width=0.85\textwidth]{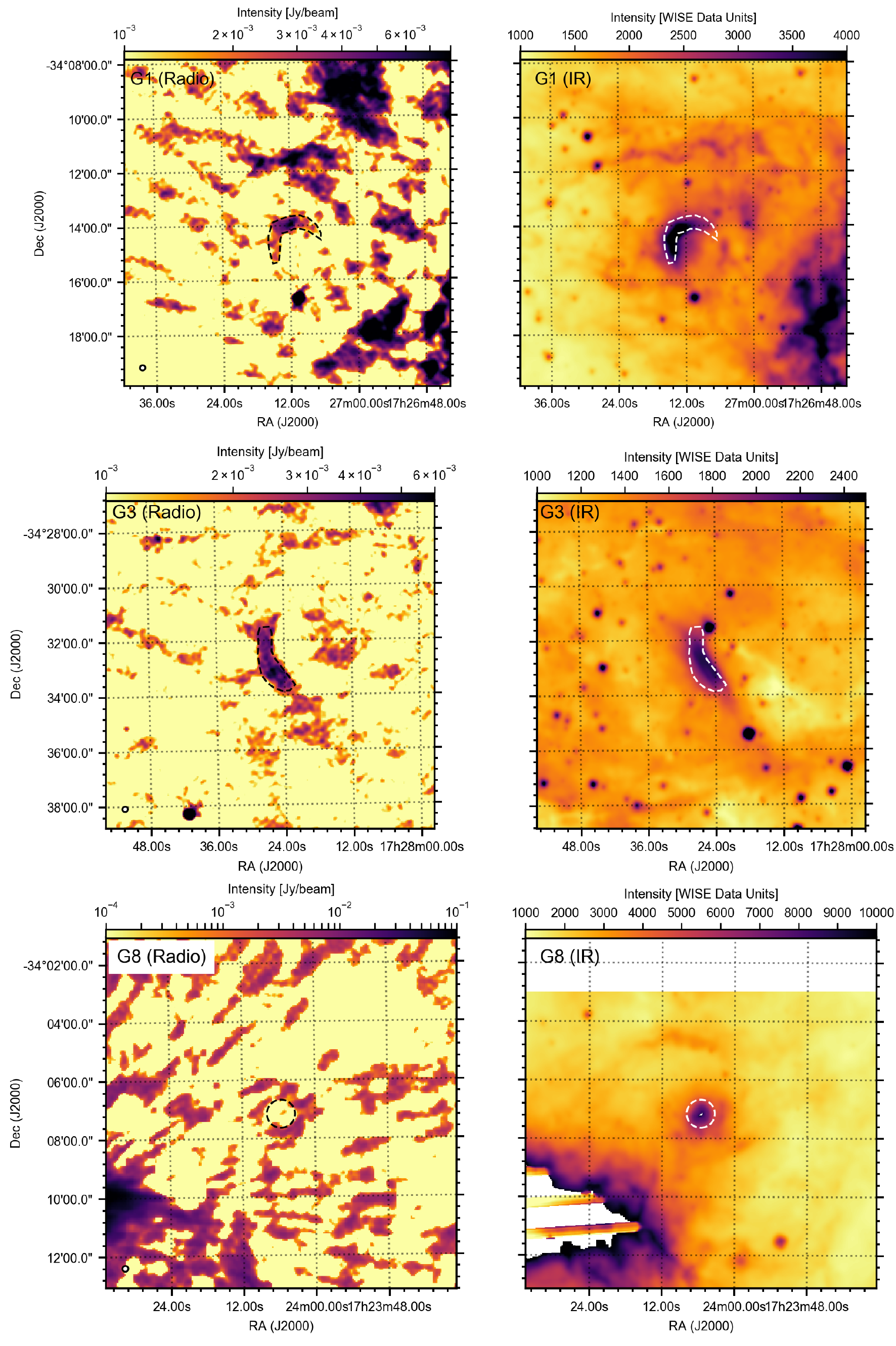}
 \caption{Radio (RACS; left column) and IR (WISE; right column) images of G1 (top), G3 (middle), and G8 (bottom). The physical scale is the same between the radio and IR images. In both images, the same contours are drawn, based on the radio morphology (G1/G3) or known position of the source (G8). We re-iterate that the contours are intended only to guide the eye in the comparison.}
 \label{fig:Gfield_radvswise}
\end{figure*}

In Figure \ref{fig:Gfield}, we show a large-scale region as observed in RACS covering NGC 6357, a star-forming region \citep{gvaramadze2011} that contains eight IR bow-shock candidates in the E-BOSS catalogues. For three of those IR bow-shocks -- G1, G3, and G8 in the terminology of \citet{peri2015} -- we show an inset zoom of the $0.2\times0.2$ degree square surrounding their locations, with their radio flux density plotted on a log scale. This scale is different from the main image and tailored for each zoomed inset to show the relevant features; the scale bar is shown for each respective field in the left column of Figure \ref{fig:Gfield_radvswise}. The insets for G1 and G3 show extended emission with an arc-like morphology, reminiscent of a typical bow shock shape, that we indicate with the black dashed lines to guide the eye. We also show the G8 inset for contrast, as the IR bow shock's location highlighted by the black dashed circle does not contain any hint of a radio counterpart. 

The left column of Figure \ref{fig:Gfield_radvswise} shows the same insets as shown in Figure \ref{fig:Gfield}, now including coordinate definitions and colorbars; for each field, the associated right panel shows its WISE Band W3 IR counterpart. The regions plotted in the left panel, defined based on the radio morphology for G1 and G3, are also shown as white dashed regions in the right panels. The fields of view are also matched. It is clear how, for G1 and G3, the extended radio and IR sources overlap in position and are alike in their bow shock-like morphology, although small differences in exact structure appear to be present. The IR source identified as the G8 bow shock in E-BOSS, however, clearly has no associated radio counterpart. Similarly, none of the remaining 5 bow shocks in this field are radio-detected. Therefore, we conclude that the G1 and G3 radio counterparts are most likely radio detections of the bow shocks. We do however stress that the NGC 6357 region is complicated, both in IR and radio, and especially for G1, many other diffuse structures are present in the field. We will encounter similar issues in the other fields below. 

\begin{figure}
 \includegraphics[width=\columnwidth]{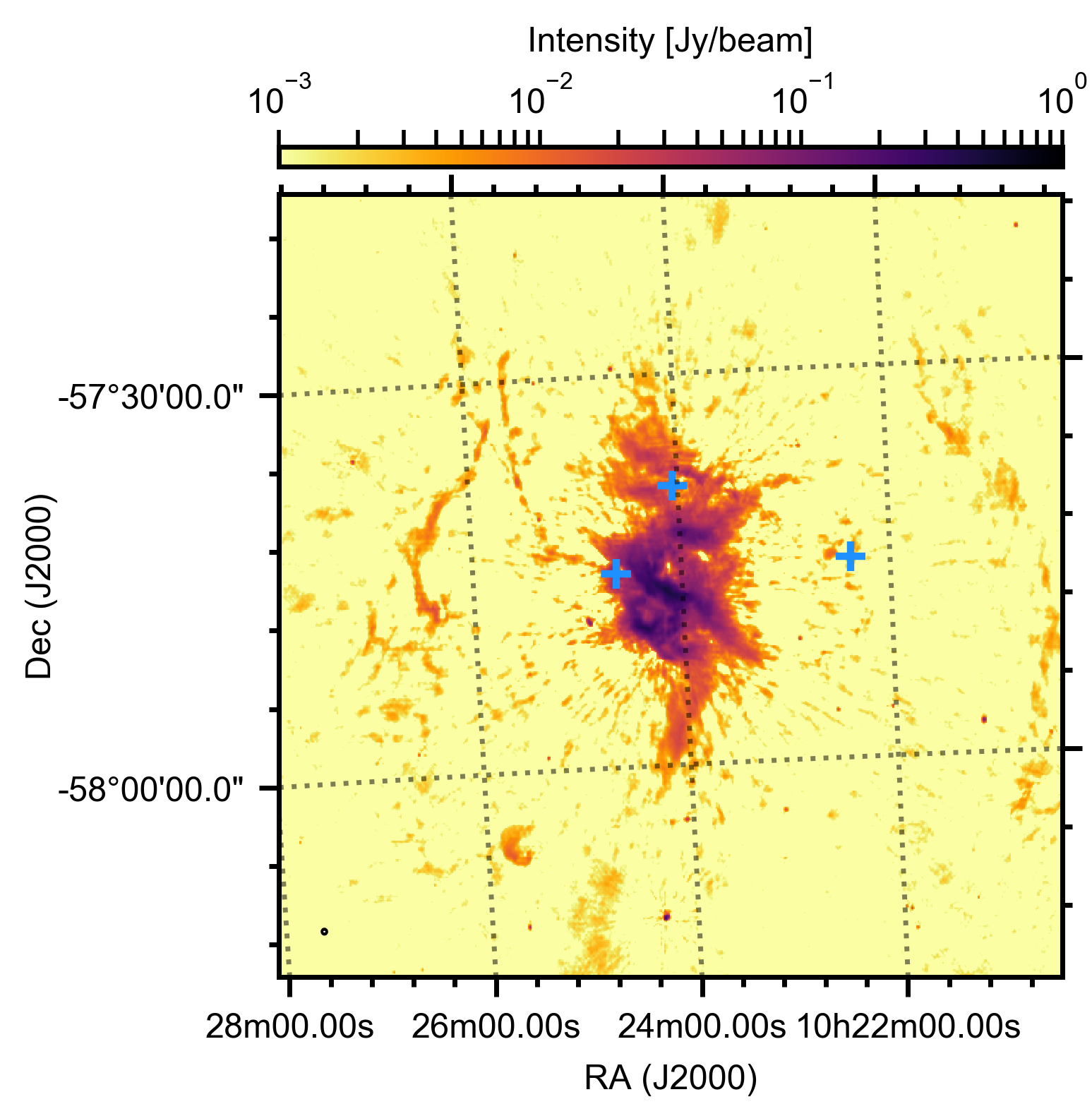}
 \caption{Logaritmically-scaled RACS image of the star-forming region RCW 49. Similar to Figure \ref{fig:Gfield}, we show the position of three known bow shocks, S1, S2, and S3. 
 Zooms of these positions are shown in Figure \ref{fig:RCW49field_radvswise}.}
 \label{fig:RCW49_field}
\end{figure}

In a similar fashion to NGC 6357, we show the massive star-forming region RCW 49 in Figure \ref{fig:RCW49_field}. \citet{peri2015} include three bow-shock candidates -- S1, S2, and S3 -- in this region, based on the \textit{Spitzer} GLIMPSE images reported by \citet{povich2008}, although in WISE data, only S1 is not saturated. Earlier, \citet{benaglia2013} studied a deep ATCA $5.5$ GHz image of RCW 49, highlighting radio emission coincident with the bow shock positions of S1 and S3 (for S1, no 9 GHz image was available; at the positon of S3, no significant 9-GHz emission was identified). We again show zoomed insets of the $0.2\times0.2$ degree squares around the positions of the three IR bow shock candidates. We see that in RACS, we encountered similar issues as \citet{peri2015} do with WISE. S1, located significantly offset from the center of the star-forming region, appears to show a radio counterpart. However, S2 and S3 are located close to the star-forming region's center, which makes it challenging to assess the origin of their radio emission.

\begin{figure*}
 \includegraphics[width=\textwidth]{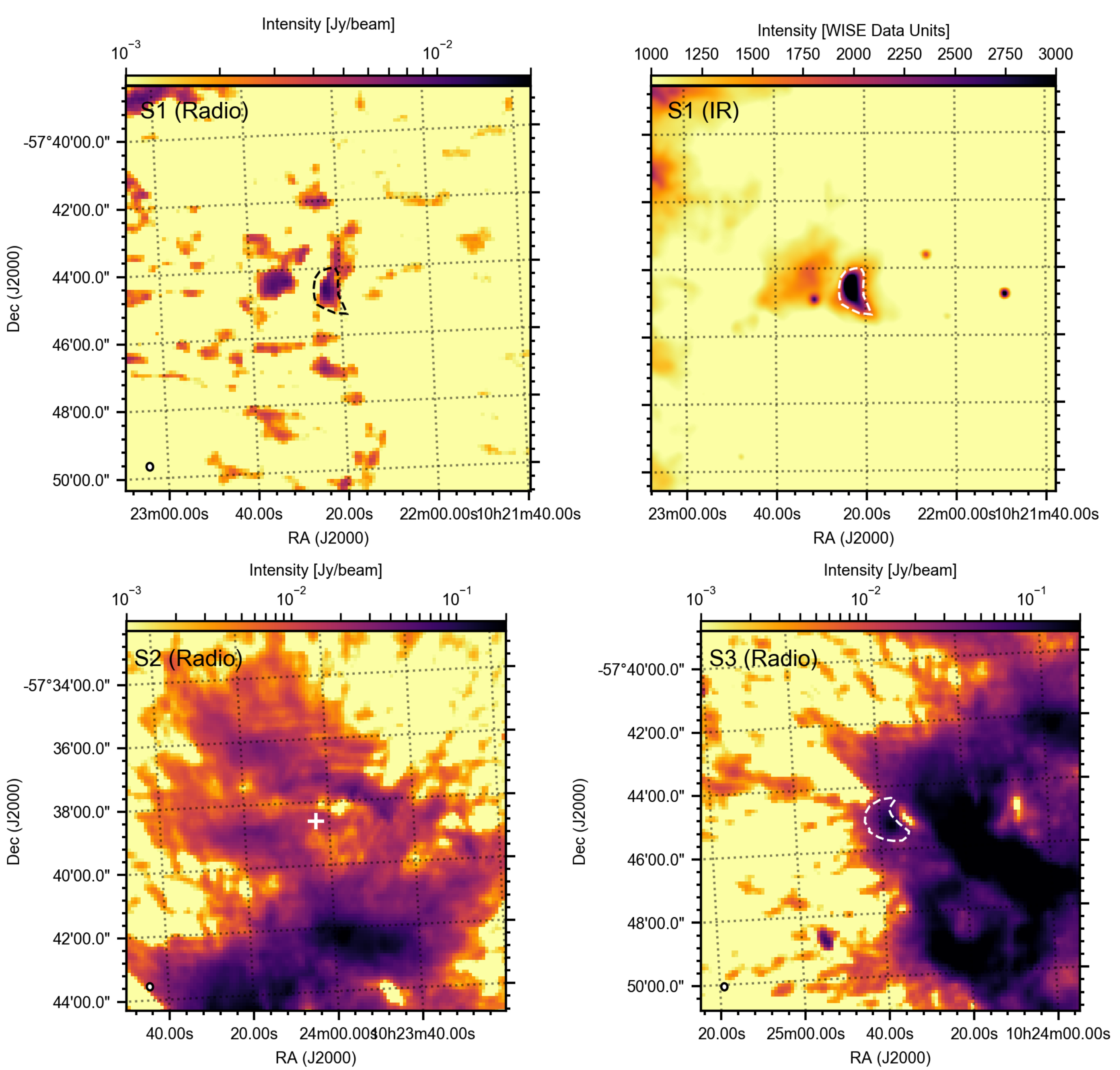}
 \caption{The top row shows the radio (RACS; left) and IR (WISE; right) images of S1 in RCW-49. Both images share the same physical scale and contour, which is drawn to guide the eye based on the IR image. The bottom row shows the radio images of S2 and S3. No un-saturated IR images from WISE are available for these bow shocks. The contour for S3 is tentatively drawn based on the radio image. We stress that the bow shock shape of S3 is not known from other wavelengths, and therefore this tentative contour is very speculative. Again, all contours are drawn only to guide the eye. Similarly, no bow shock shape is known for S2 from other wavelengths.}
 \label{fig:RCW49field_radvswise}
\end{figure*}

In Figure \ref{fig:RCW49field_radvswise}, we show S1 in the radio (left) and IR (right) bands in the top panels. The drawn region is, in these images, based on the IR bow shock, showing how the radio source aligns with the IR structure. Therefore, we identify the RACS source as the radio counterpart of the S1 IR bow shock, noting that the radio source extends further North. This further extension may be an unrelated extended radio source, or part of the noisy background close to the bright RCW 49 region. For S2 (left) and S3 (right), shown in radio in the bottom panels, the lack of unsaturated WISE data makes it challenging to assess the presence of a radio counterpart. 
For S3, an apparent gap to the West of the IR bow shock location of S3 can be seen. Therefore, we can very tentatively draw an arc-shaped region at this position, but to identify this as a radio bow shock remains extremely speculative. For S2, indicated by the white cross, a decrease in radio flux can be seen to the South of the bow shock position. However, this decrease is less pronounced than for S3. All combined, we can therefore conclude that radio emission is seen at the position of S2 and S3, but it could conceivably be unrelated to the bow shock. 

\begin{figure*}
 \includegraphics[width=\textwidth]{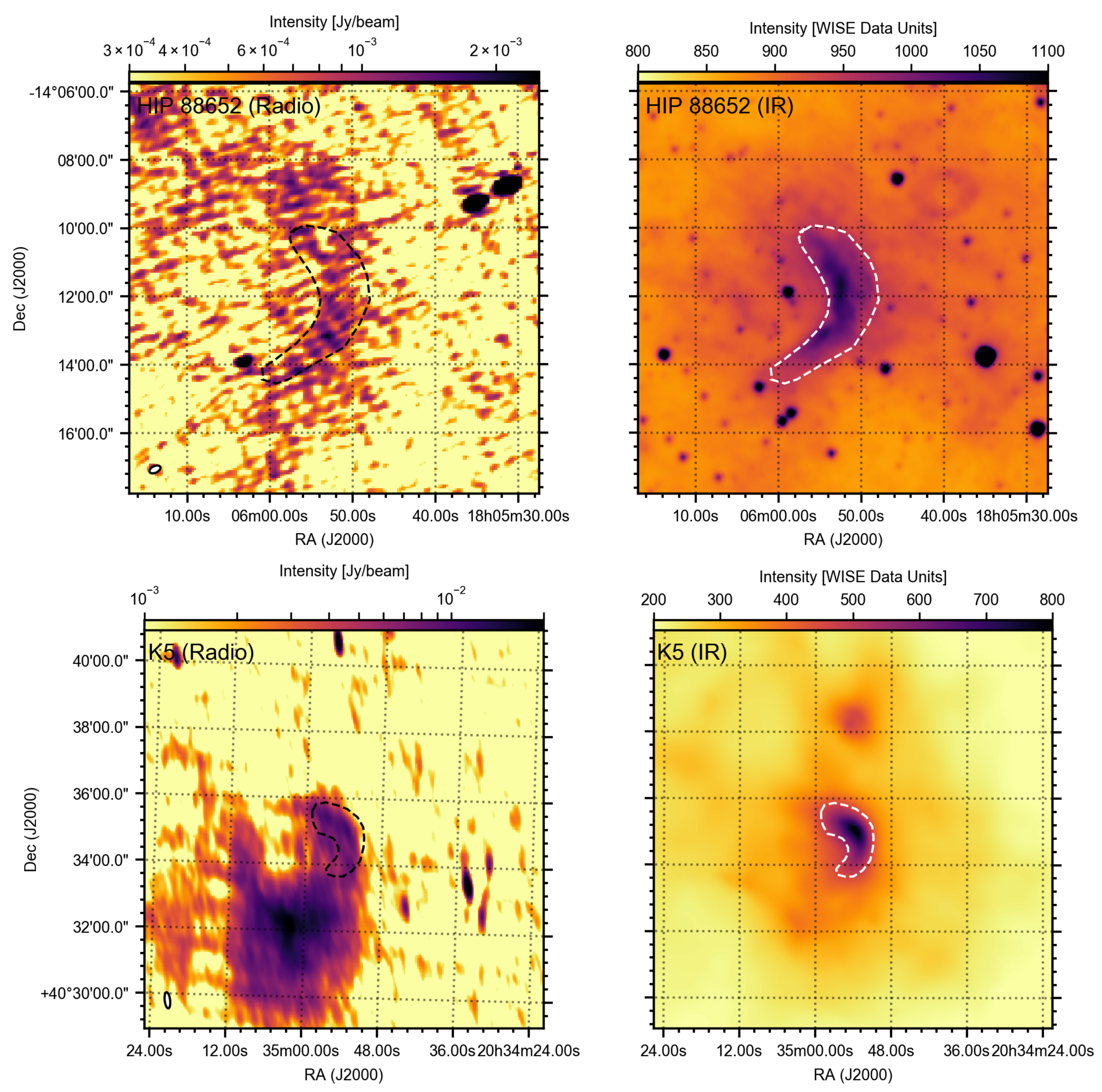}
 \caption{Radio (RACS; left column) and IR (WISE; right column) images of HIP 88652 (top) and K5 (bottom). The physical scale and contours are the same between the radio and IR images. The contours, in both rows, are drawn based on the IR image and are intended to guide the eye in comparing the two wavebands.}
 \label{fig:HIP88652_K5_field_radvswise}
\end{figure*}

\begin{figure*}
 \includegraphics[width=\textwidth]{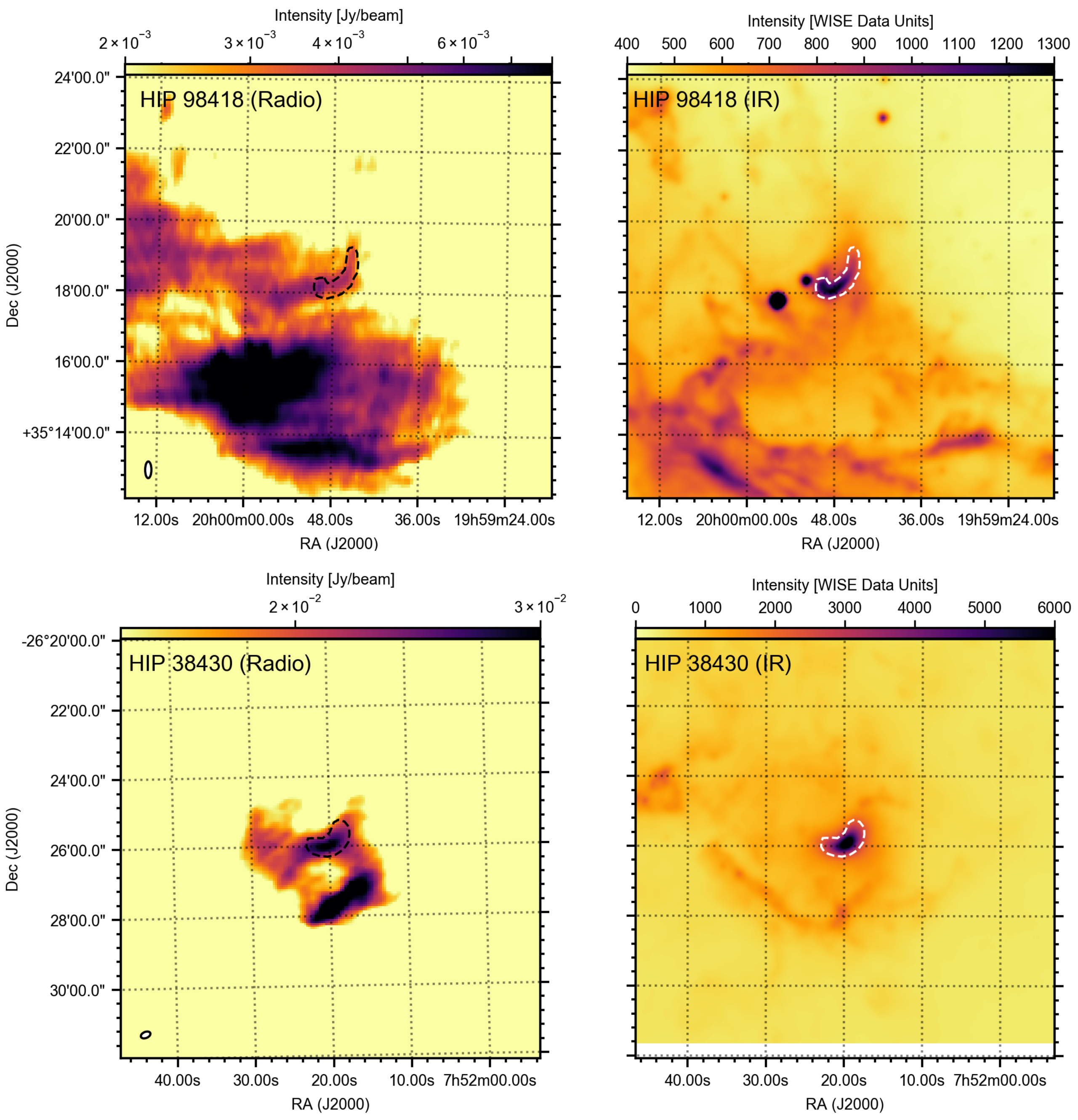}
 \caption{Radio (RACS; left column) and IR (WISE; right column) images of HIP 98418 (top) and HIP 38430 (bottom). The physical scale and contours are the same between the radio and IR images. The contours in both rows are drawn based on the IR shape of the bow shock; all contours are intended to guide the eye in comparing the two wavebands.}
 \label{fig:HIP98418_38430field_radvswise}
\end{figure*}

The next four IR bow shocks with radio emission at their position are shown in Figures \ref{fig:HIP88652_K5_field_radvswise} and \ref{fig:HIP98418_38430field_radvswise}, in the same fashion as Figure \ref{fig:Gfield_radvswise}. In all cases, the black and white dashed regions are the same in the left and right columns; in both these figures, all four regions are based on the morphology of the IR bow shock in the right column. The first of the four fields, centered on HIP 88652 (Figure \ref{fig:HIP88652_K5_field_radvswise}, top) shows a large, extended radio source at the position of the IR bow shock. The brightest region of this extended source appears to follow the arc-like IR shape, although the radio structure extends in both the North and South directions. The radio image is also heavily affected by the presence of two bright point sources to the North-West of the bow shock, that remain incompletely de-convolved and therefore imprint radial streaks of radio residuals that intersect with the bow shock position. Therefore, while the radio morphology and positional overlap suggest that the RACS structure may be the radio counterpart of the HIP 88562 bow shock, this conclusion requires confirmation; either by new radio observations or possibly by re-imaging the ASKAP data tailored specifically for this field.

The bottom panels of Figure \ref{fig:HIP88652_K5_field_radvswise} show the source called K5 in the second E-BOSS catalogue, located in the Cygnus OB2 association. This association contains a total of 11 bow shocks listed in E-BOSS, including the first radio stellar runaway bow shock BD+43$^{\rm o}$3654 (which is too far north to be covered by RACS). Of the covered bow shocks, only K5 shows radio emission in RACS. The black dashed region, constructed based on the WISE image, shows how the radio emission broadly traces the bow shock, merging into the extended radio source to the South. Despite the presence of this Southern extended source, the similarity in morphology makes us conclude that the radio emission likely forms the RACS counterpart of K5. 

The two fields shown in the radio (left) and IR (right) bands in Figure \ref{fig:HIP98418_38430field_radvswise} are similar to the case of K5: the regions, based on the IR bow shock morphology, show how radio emission traces the IR morphology. However, for both HIP 98418 (top) and HIP 38430 (bottom), other extended radio emission is located close to the bow shock location. In both cases, the target is located in a complex ISM field, as is visible from the IR images and confirmed with RACS. However, similar to K5, the similarity in morphology and position leads us to conclude that the emission is likely due to the radio counterpart of the IR bow shock.

\begin{figure*}
 \includegraphics[width=\textwidth]{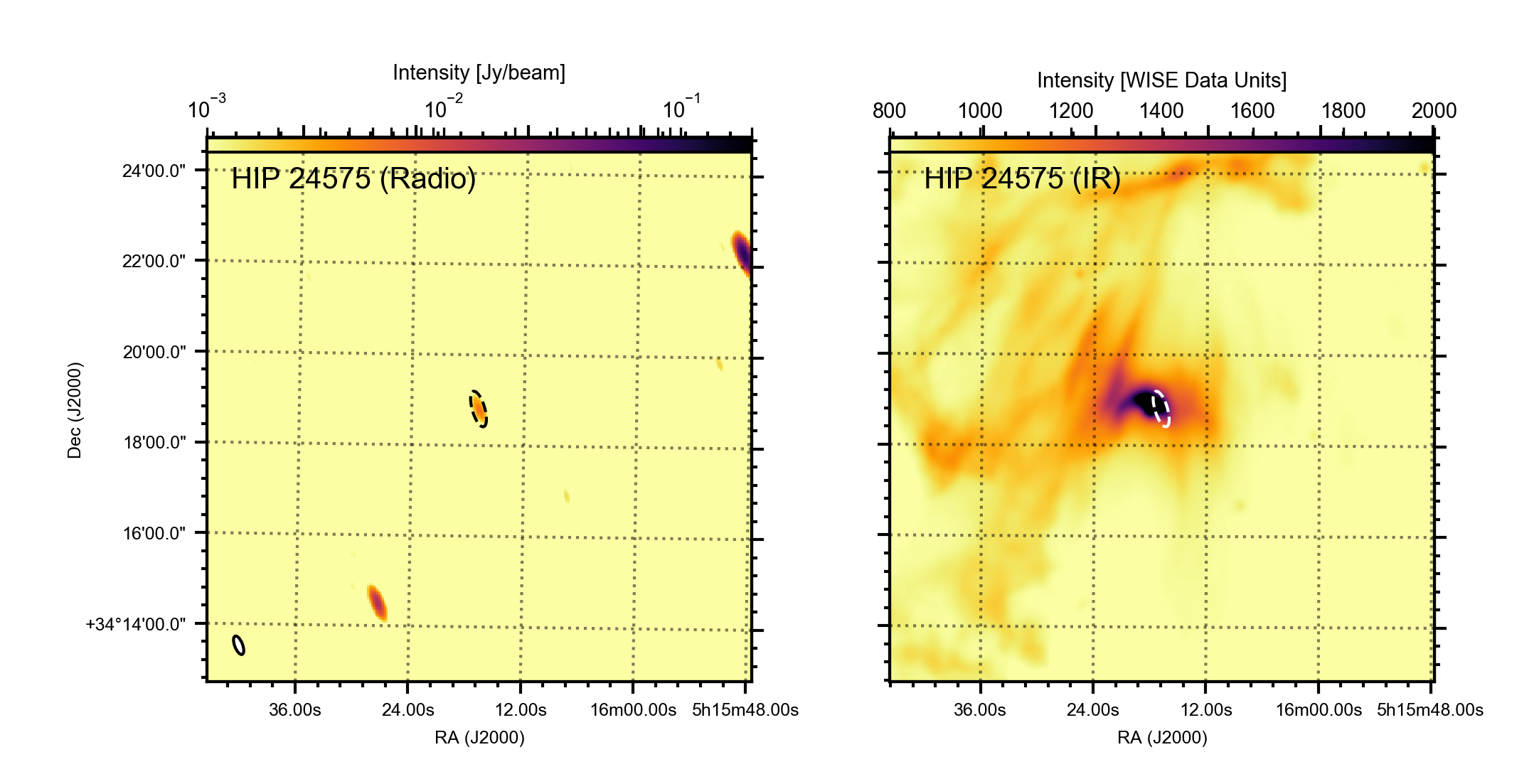}
 \caption{Radio (RACS; left) and IR (WISE; right) images of HIP 24575. The physical scale and contours are the same between the radio and IR images. The contours are based on the radio source and intended to guide the eye in comparing the two wavebands. The radio source at the position of HIP 24575 appears to be consistent with a point source.}
 \label{fig:HIP24575_field_radvswise}
\end{figure*}

The complexity of the radio fields, seen in the targets above, makes it challenging to assess beyond doubt whether detected radio emission originates in the bow shock. In the tenth and final E-BOSS target with associated radio emission, we encounter the simplest field in RACS. In Figure \ref{fig:HIP24575_field_radvswise}, we show the RACS and WISE field containing HIP 24575 (i.e. the star in AE Aur). The region plotted in both panels is based on the shape of the radio source, which is consistent with a point source. While it overlaps with the position of the IR bow shock of HIP 24575, the radio image does not suggest we are looking at a radio bow shock. Especially its point source nature, without any hints for even faint extended emission surrounding it, argues for a coincident interloper. In addition, VLA radio studies at higher radio frequencies also did not detect any radio emission at the position of the bow shock, down to sensitivities below the flux density of the RACS source \citep{rangelov2019}. Based on the stellar wind properties and distance of HIP 24575 as listed in \citet{peri2015}, we can use the formalism in \citet{wright1975} to estimate the stellar wind's radio luminosity. We expect a sub-nJy flux density from the massive star's wind, which cannot be reconciled with the RACS point source flux density in the mJy range. Therefore, we conclude that this radio source is likely unrelated to either the massive star HIP 24575 or its bow shock. 

\section{Measurements and calculations}
\label{sec:calcs}

Having assessed the presence and morphology of radio emission at the position of the E-BOSS bow shocks, we can turn to estimating their basic observables and physical properties. We will do so for both radio-detected and non-detected targets, but do exclude a number of bow shocks: firstly, we do not consider Vela X-1, as RACS does not reveal radio emission at its position, while a MeerKAT radio detection is presented in detail in \citet{vandeneijnden2021}. Secondly, the sources referred to as SER1 to SER7 in \citet{peri2015} do not have an associated distance in the catalogue. Therefore, as none of these sources are radio detected and the lack of distance prevents further physically-meaningful constraints to be made, we will ignore these targets in the remainder of this work. Finally, we do not consider RCW 49 S2 and RCW 49 S3 here, as the E-BOSS catalogue does not contain geometrical IR constraints for these two targets. While both do show radio emission at their position \citep[see also][]{benaglia2013}, the connection between this emission and the bow shock also remains unclear and requires dedicated follow-up observations.

\begin{table*}
\caption{Summary of the radio measurements for the bow shocks considered in the calculations in Section \ref{sec:calcs}. *To aid the reproduction of our calculations, we also list the mass-loss rate and wind velocity, the ISM density, and the source distance, as listed in the E-BOSS catalogue \citep{peri2012,peri2015}, as well as the RACS field. Note that, in addition to these 40 sources, we searched for radio emission from 10 more targets (Vela X-1, SER 1--7, and S2 and S3 in RCW 49). As discussed in the text, these are not considered in the calculations for various reasons. The geometrical bow shock properties can be found in the E-BOSS catalogue \citep{peri2012,peri2015}.}
\label{tab:measurements}
\begin{tabular}{lllllllllll}
\hline
Source & Detection & $S_{\nu}$ or RMS & Image RMS & Beam major & Beam minor & $\dot{M}_{\rm wind}$* & $v_{\rm wind}$* & D* & $n_{\rm ism}$* & RACS field \\
& & [$\mu$Jy] & [$\mu$Jy] & axis ["] & axis ["] & [$\rm M_{\odot}$/yr] & [km/s] & [pc] & [cm$^{-3}$] & \\ 
\hline
HIP 16518 & No & 254 & 285 & 19.9 & 17.7 & $6\times10^{-9}$ & $500$ & $650$ & $0.2$ & 0334-37A\\ 
HIP 24575 & No & 680 & 245 & 19.8 & 16.8 & $1\times10^{-7}$ & $1200$ & $548$ & $3$ & 0510-31A\\ 
HIP 25923 & No & 213 & 267 & 19.8 & 15.9 & $6\times10^{-8}$ & $1000$ & $900$ & $1$ & 0535-06A\\ 
HIP 26397 & No & 250 & 282 & 18.7 & 15.4 & $1.4\times10^{-8}$ & $750$ & $350$ & $2$ & 0537-37A\\ 
HIP 29276 & No & 230 & 228 & 19.9 & 16.0 & $1\times10^{-9}$ & $600$ & $400$ & $0.003$ & 0559-56A\\ 
HIP 31766 & No & 270 & 304 & 24.7 & 16.3 & $1.07\times10^{-6}$ & $1590$ & $1414$ & $0.03$ & 0649+00A\\ 
HIP 32067 & No & 1330 & 203 & 14.2 & 12.7 & $1.3\times10^{-7}$ & $2960$ & $2117$ & $0.1$ & 0649-06A\\ 
HIP 34536 & No & 760 & 733 & 14.6 & 12.6 & $1.9\times10^{-7}$ & $2456$ & $1293$ & $0.01$ & 0709-12A\\ 
HIP 38430 & Yes & 31700 & 268 & 21.6 & 15.6 & $7\times10^{-7}$ & $2570$ & $900$ & $60$ & 0741-25A\\ 
HIP 62322 & No & 180 & 213 & 20.1 & 13.4 & $6\times10^{-9}$ & $300$ & $150$ & $0.02$ & 1230-68A\\ 
HIP 72510 & No & 250 & 226 & 18.3 & 13.4 & $2.7\times10^{-7}$ & $2545$ & $350$ & $0.2$ & 1428-56A\\ 
HIP 75095 & No & 250 & 281 & 18.7 & 13.2 & $1.4\times10^{-7}$ & $1065$ & $800$ & $40$ & 1510-56A\\ 
HIP 77391 & No & 200 & 277 & 15.6 & 13.1 & $2.5\times10^{-7}$ & $1990$ & $800$ & $30$ & 1600-50A\\ 
HIP 78401 & No & 330 & 272 & 17.7 & 17.3 & $1.4\times10^{-7}$ & $1100$ & $224$ & $2$ & 1550-25A\\ 
HIP 81377 & No & 270 & 233 & 13.9 & 12.8 & $2\times10^{-8}$ & $1500$ & $222$ & $1$ & 1625-12A\\ 
HIP 82171 & No & 170 & 224 & 18.3 & 12.7 & $9\times10^{-8}$ & $1345$ & $845$ & $1$ & 1635-56A\\ 
HIP 88652 & Yes & 2240 & 327 & 22.1 & 15.9 & $5\times10^{-8}$ & $1535$ & $650$ & $2$ & 1806-12A\\ 
HIP 92865 & No & 310 & 277 & 15.3 & 12.8 & $4\times10^{-8}$ & $1755$ & $350$ & $0.003$ & 1849-06A\\ 
HIP 97796 & No & 410 & 288 & 23.5 & 14.6 & $5\times10^{-8}$ & $1980$ & $2200$ & $0.02$ & 1938-18A\\ 
HIP 47868 & No & 210 & 239 & 18.9 & 13.2 & $3\times10^{-8}$ & $1200$ & $1075.3$ & $2.6$ & 0952-31A\\ 
HIP 98418 & Yes & 5300 & 263 & 18.8 & 12.9 & $2.4\times10^{-8}$ & $2545$ & $529.1$ & $380$ & 1954-37A\\ 
HIP 104579 & No & 340 & 188 & 14.8 & 14.0 & $3\times10^{-8}$ & $650$ & $1149.4$ & $0.7$ & 2056-37A\\ 
HD 57682 & No & 430 & 285 & 23.9 & 14.7 & $1.6\times10^{-8}$ & $1900$ & $1600$ & $85$ & 0709-12A\\ 
HIP 86768 & No & 290 & 283 & 24.1 & 14.4 & $3\times10^{-8}$ & $550$ & $737$ & $0.1$ & 1735-06A\\ 
RCW 49 S1 & Yes & 12600 & 229 & 15.2 & 13.1 & $3.23\times10^{-6}$ & $2800$ & $6100$ & $30$ & 1014-56A\\ 
K4 & No & 1900 & 187 & 15.1 & 13.8 & $3\times10^{-8}$ & $300$ & $1500$ & $2.8$ & 2025-37A\\ 
K5 & Yes & 11150 & 187 & 15.1 & 13.8 & $5\times10^{-8}$ & $1500$ & $1500$ & $2$ & 2025-37A\\ 
K7 & No & 1610 & 187 & 15.1 & 13.8 & $1.5\times10^{-6}$ & $2500$ & $1500$ & $44$ & 2025-37A\\ 
K10 & No & 530 & 187 & 15.1 & 13.8 & $3\times10^{-8}$ & $550$ & $1500$ & $1.8$ & 2025-37A\\ 
G1 & Yes & 8000 & 376 & 13.6 & 12.8 & $2\times10^{-7}$ & $2100$ & $1700$ & $14$ & 1721-37A\\ 
G2 & No & 1500 & 376 & 13.6 & 12.8 & $4\times10^{-7}$ & $2250$ & $1700$ & $14$ & 1721-37A\\ 
G3 & Yes & 5970 & 376 & 13.6 & 12.8 & $4\times10^{-7}$ & $2000$ & $1700$ & $16$ & 1721-37A\\ 
G4 & No & 780 & 376 & 13.6 & 12.8 & $5\times10^{-7}$ & $2550$ & $1700$ & $42$ & 1721-37A\\ 
G5 & No & 1700 & 376 & 13.6 & 12.8 & $1\times10^{-7}$ & $2000$ & $1700$ & $4$ & 1721-37A\\ 
G6 & No & 1600 & 376 & 13.6 & 12.8 & $1\times10^{-7}$ & $1000$ & $1700$ & $13$ & 1721-37A \\ 
G8 & No & 4500 & 376 & 13.6 & 12.8 & $4\times10^{-8}$ & $1500$ & $1700$ & $5$ & 1721-37A\\ 
4U 1907+09 & No & 600 & 271 & 20.0 & 14.4 & $7\times10^{-7}$ & $2900$ & $4000$ & $0.1$ & 1922-12A\\ 
J1117-6120 & No & 2640 & 250 & 18.0 & 13.0 & $6\times10^{-7}$ & $2600$ & $7600$ & $6.4$ & 1135-62A\\ 
BD 14-5040 & No & 730 & 527 & 21.5 & 14.8 & $3\times10^{-8}$ & $400$ & $1800$ & $0.1$ & 1831-12A\\
Star 1 & No & 1900 & 527 & 21.5 & 14.8 & $6.3\times10^{-7}$ & $2200$ & $1800$ & -- & 1831-12A\\ 
\hline
\end{tabular}\\
\end{table*}

In Table \ref{tab:measurements}, we list the relevant observables for the 40 remaining targets. For each source, we either list the flux density detected in its brightest beam (detected sources) or the RMS measured over a circular region with a 2 arcminute radius centered on the bow shock position (non-detected sources). We also list the full image RMS (column 4), as reported in the RACS data repository, which can be used as an RMS estimate for the radio detected sources. Alternatively, comparing these image RMSs with the RMS at the source position highlights how for several non-detected targets, locally-enhanced noise may have complicated the detection of a radio counterpart. For the detected targets, we only consider the brightest beam, because the bow shock shape is not as well defined as for their IR counterparts or because close-by confusing sources complicate the determination of the region where the bow shock dominates. Therefore, assuming the brightest beam is most clearly dominated by bow shock emission, we will restrict our calculations in this section to that part of the shock. Any physical scenario that can explain the observed emission should be able to explain the most extreme part of the bow shock; therefore, the brightest beam will be most constraining in assessing physical explanations. We also list the beam's major and minor axis size for each image, as well as the RACS image name for easy data accessibility.

These simple observables, in combination with the geometrical IR measurements, distances, stellar wind properties, and ISM density estimates listed in the E-BOSS catalogues (and partially tabulated in the Appendix as well), allow us to explore two scenarios for the radio emission. As introduced extensively in \citet{vandeneijnden2021} to investigate the radio emission from the bow shock of Vela X-1, we can consider non-thermal/synchrotron or thermal/free-free emission processes -- as well as, naturally, their combination. We summarize the outcome of these considerations in Table \ref{tab:scenarios}. The calculations discussed below can be reproduced using the Jupyter notebook accompanying this work (see Data Availability).

\begin{table*}
\caption{Summary of the thermal and non-thermal calculations for the seven (candidate) radio bow shocks with geometrical constraints (i.e. all except S2 and S3 in RCW 49). For each source, we list whether synchrotron or free-free emission can account for the observed radio source, with the main argument for this assessment. In the final column, we summarize the conclusion regarding the nature of the emission. *Also shows an H$\alpha$ counterpart consistent with a thermal scenario.}
\label{tab:scenarios}
\begin{tabular}{llllll}\hline
Source / & Synchrotron & Synchrotron & Free-free & Free-free & Conclusion \\ 
field & origin & constraints & origin & constraints & \\ \hline
G1 / NGC 6357 & No & $\eta_e > 100$\% for all $B$ & Yes & $n_e(T; S_\nu)/n_{\rm ISM}$ between $4$--$10$ for realistic $T$ & Thermal \\
G3 / NGC 6357 & Unlikely & $\eta_e \gtrsim 40$\% for all $B$ & Yes* & $n_e(T; S_\nu)/n_{\rm ISM}$ between $4$--$10$ for realistic $T$ & Thermal \\ 
S1 / RCW 49 & Unlikely & $\eta_e \gtrsim 50$\% for all $B$ & Yes & $n_e(T; S_\nu)/n_{\rm ISM}$ between $4$--$10$ for realistic $T$ & Thermal\\
HIP 88652 & Yes & $\eta_e < 10$\% for $B\gtrsim50$ $\mu$G & No & $n_e(T; S_\nu)/n_{\rm ISM} \gg 10$ & Non-thermal \\
K5 / Cyg OB2 & No & $\eta_e > 100$\% for all $B$ & No & $n_e(T; S_\nu)/n_{\rm ISM} \gg 10$ & Neither \\
HIP 98418 & Yes & $\eta_e < 10$\% for $B\gtrsim30$ $\mu$G & Maybe & Requires high $T$ or an overestimated $n_{\rm ISM}$ & Non-thermal \\
HIP 38430 & Yes & $\eta_e < 10$\% for $B\gtrsim60$ $\mu$G & Yes & $n_e(T; S_\nu)/n_{\rm ISM}$ between $4$--$10$ for realistic $T$ & Either/combined \\ \hline
\end{tabular}\\
\end{table*}

\subsection{Non-thermal/synchrotron emission}

Here, we first assess a synchrotron origin. In this scenario, developed over the past decade by many authors \citep[e.g.][]{benaglia2010, benaglia2021, delvalle2012, delvalle2018, debecker2017, delpalacio2018}, a fraction of the kinetic power of the stellar wind powers the acceleration of electrons at the shock into a power-law number density distribution, likely via diffusive shock acceleration. These electrons then gyrate around the magnetic field, originating in the stellar wind but possibly amplified in the shock, therefore losing energy via synchrotron radiation. Importantly, several fundamental parameters cannot be determined directly from the RACS data but can only be estimated: the magnetic field strength $B$ in the shock, the maximum energy of the electrons $E_{\rm max}$, and the slope of the relativistic electron number density distribution $p$ (related to the radio spectral index $\alpha$ via $p = 2\alpha+1$, where $S_\nu \propto \nu^{-\alpha}$). Similarly, the efficiency of injecting energy into the relativistic electron population from the stellar wind's kinetic power -- the fundamental energy source of the electrons in this scenario -- is not known nor uniquely measurable. However, for reasonable assumptions of these parameters, we can consider whether a realistic and self-consistent scenario can be constructed, fitting with the observed radio properties.

Here, we will pay particular attention to the aforementioned injection efficiency $\eta_e$: the fraction the total available kinetic power of the stellar wind passing through the considered bow shock region required to maintain the electron population in a steady state. For assumed values of $p$ and $E_{\rm max}$, and assuming that the energy loss of electrons is driven by diffusive over radiative losses\footnote{Thereby, formally, yielding a lower limit on $\eta_e$: extra loss mechanisms naturally increase the required energy and therefore efficiency to maintain the steady state.}, one can derive the following relation between $\eta_e$ and $B$ (\citealt{vandeneijnden2021}, but see also e.g. \citealt{delpalacio2018} and \citealt{benaglia2021} for equivalent calculations):

\begin{equation}
    \begin{split}
    \eta_e \approx &\frac{128\pi^3 R_0^3 D^2 S_\nu \epsilon_0 c m_e}{3\sqrt(3)\dot{M}_{\rm wind} v_{\infty} \Delta V_{\rm bowshock} e^3 B a(p)}\\
    &\times \left(\frac{3eB}{2\pi\nu m_e^3 c^4}\right)^{-(p-1)/2} \int_{E_{\rm min}}^{E_{\rm max}} E^{1-p}dE \text{ .}
    \end{split}
    \label{eq:eta}
\end{equation}

\noindent In the above Equation, $R_0$ is the standoff distance (i.e. the distance between the bow shock apex and the star); $D$ is the distance to the source; $S_{\nu}$ is the flux density of the considered region of the bow shock (i.e. integrated over the entire structure or a subset, such as the brightest beam); $\dot{M}_{\rm wind}$ is the stellar wind mass-loss rate; $v_{\infty}$ is the terminal wind velocity; $\Delta$ is the bow shock width (assumed to equal its depth in the derivation of this Equation); $V_{\rm bowshock}$ is the volume of the considered bow shock region; $B$ is the magnetic field; $a(p)$ is a numerical function, defined in Equation 8.129 of Longair (2011); $\nu$ is the observing frequency; $E_{\rm min}$ is the minimum electron energy, here assumed to be the electron rest mass; and finally, $\epsilon_0$, $c$, $m_e$, and $e$ are the vacuum permittivity, speed of light, electron mass, and unit charge, respectively. 

The injection efficiency as a function of magnetic field is conveniently bounded by two simple conditions. Firstly, by definition, the efficiency cannot exceed one. Secondly, the magnetic field cannot exceed a maximum value $B_{\rm max}$, imposed by the condition that the stellar wind should be compressible in order for the bow shock to have formed \citep{delpalacio2018,benaglia2021}:

\begin{equation}
\frac{B_{\rm max}^2}{8\pi} = \frac{2}{1+\gamma_{\rm ad}} \rho_{\rm wind} v^2_{\infty} \text{ ,}
\end{equation}

\noindent where $\gamma_{\rm ad} = 5/3$ is the adiabatic coefficient for an ideal gas and $\rho_{\rm wind}$ is the stellar wind density at $R_0$. Therefore, $\eta_e$ shows a monotonically decreasing dependence\footnote{Since $\eta_e \propto B^{-(p+1)/2}$ and $p > 0$ for a realistic electron population accelerated at a shock.} on magnetic field between $\eta_e = 1$ and its minimum at $B_{\rm max}$ -- or, alternatively, $\eta_e(B_{\rm max})$ exceeds one and the synchrotron scenario is unfeasible as the only emission mechanism. Importantly, an additional and more stringent constraint on $\eta_e$ follows from observations of diffusive shock acceleration in various contexts. As discussed in detail in \citet{vandeneijnden2021}, observations of the BD+43$^{\rm o}$3654 radio bow shock \citep{delpalacio2018,benaglia2021}, pulsar wind nebulae \citep{stappers2003}, and colliding wind binaries \citep{delpalacio2021} imply typical values below $10$\%. We will, in the remainder of this paper, use this value of $\eta_e = 0.1$ as a more realistic maximum value. 

\begin{figure}
 \includegraphics[width=\columnwidth]{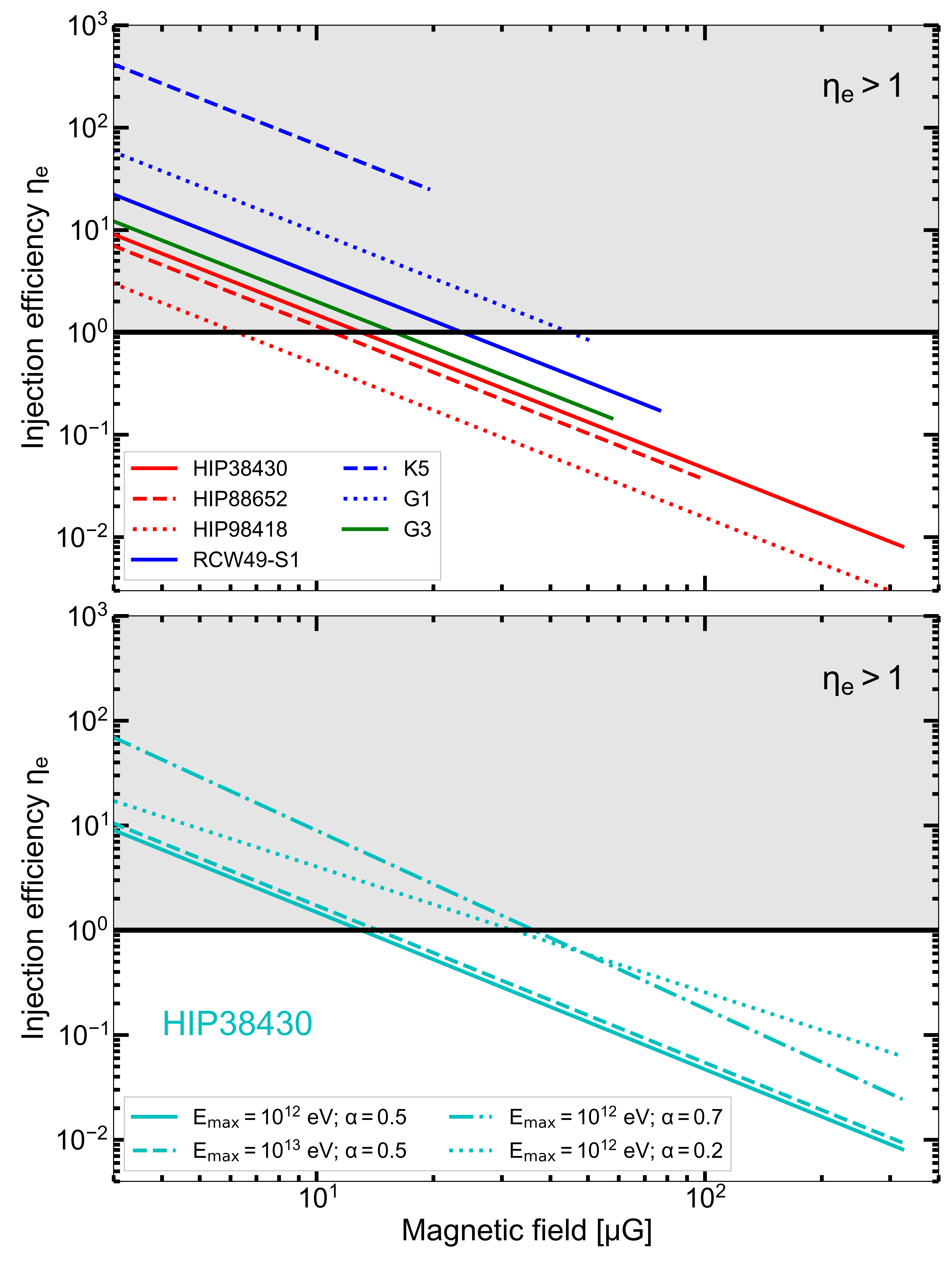}
 \caption{Top panel: the injection efficiency, as defined in Equation \ref{eq:eta}, as a function of magnetic field for the seven (candidate) radio bow shocks with known geometrical properties (i.e. excluding S2 and S3 in RCW 49). The grey shaded region indicates where the efficiency reaches above $100$\%. Only HIP 38430 and HIP 98418 reach below efficiencies of $10$\%. Bottom panel: the same relation as the top panel, plotted for HIP 38430, varying $E_{\rm max}$ and $\alpha$. A higher $E_{\rm max}$ slightly increases the required efficiencies, while changes in $\alpha$ yield larger increases.}
 \label{fig:nonthermal}
\end{figure}

In the top panel of Figure \ref{fig:nonthermal}, we plot the injection efficiency for the considered, radio-detected bow shocks. For the plotted relations, we have assumed a maximum electron energy $E_{\rm max} = 10^{12}$ eV and $p=2$ (i.e. $\alpha=0.5$). One can immediately observe how for K5 and G1, the efficiency never or only just, respectively, reaches below $100\%$, for any magnetic field strength that leaves the wind compressible. On the other hand HIP 88652, HIP 38438, and HIP 98418 reach required efficiencies below $4$\%, $1$\%, and $0.3$\%, respectively, for their maximum magnetic field strengths. Since the magnetic field strength in the stellar wind scales as one over the distance to the star, it requires significant fine tuning of the wind mass loss, velocity, and ISM density to expect that the majority of bow shocks are close to their maximum magnetic field. However, for HIP 38438 and HIP 98418, a large range of magnetic field strengths implies realistic efficiencies in the regime below $\sim 10$\%. 

That leaves two sources in between these two extremes, requiring efficiencies between $10$-$100$\% to explain the observed radio emission: G3 and RCW 49 S1. For those, we find similar results as \citet{vandeneijnden2021} reported for the Vela X-1 radio bow shock: unexpectedly high injection efficiencies (i.e. $>10$\%) are needed to assign all radio emission a non-thermal origin. We should, however, add three nuances to this statement: firstly, the non-thermal emission may instead contribute just a fraction of the emission; secondly, one should carefully consider whether this brightest radio beam is indeed representative of and dominated by emission from the bow shock; finally, such quantitative statements may depend on our assumptions regarding the electron population. We assess the final option explicitly in the lower panel of Figure \ref{fig:nonthermal}, where we plot the injection efficiency relation for HIP 38430 assuming four different combinations of $E_{\rm max}$ and $p$. Clearly, the exact numerical values depend on these assumptions, but $E_{\rm max}=10^{12}$ and $p=2$, as assumed in the top panel, imply the lowest required efficiencies. In other words, if those assumptions are inaccurate, it exacerbates the complications G3 and RCW 49 S1 in the non-thermal model. 

\subsection{Thermal/free-free emission}

In addition to a non-thermal scenario, we can instead consider a thermal/free-free origin of the emission. In such a scenario, we can consider optically-thin free-free emission, as typical bow shock densities and widths are insufficient to make the shock optically thick at radio frequencies \citep{vandeneijnden2021}. Then, the free-free emissivity and therefore observed radio flux are set by the electron temperature and density in the shock. As discussed in detail in \citet{vandeneijnden2021}, we can derive a simple scaling between electron temperature and density for a given observed radio flux; alternatively, for non-detected sources, we can instead plot electron density upper limits as a function of temperature. 

If line emission from the thermal electrons is also observed -- for instance H$\alpha$ -- the degeneracy between $n_e$ and $T$ can be broken and a unique solution can be found. However, in the absence of literature reports of such emission for the majority of bow shocks (but see Section \ref{sec:halpha}), we can instead find other constraints. For instance, we expect the post-shock temperature to scale with the shock velocity, equal to the stellar peculiar velocity $v_*$, as $kT \approx (3/16)\mu m_p v_*^2$, where $\mu \approx 0.6$ for cosmic abundances and $m_p$ is the proton mass \citep{helder2009}. For typical runaway stellar velocities of $30$ km/s and $50$ km/s, this relation implies electron temperatures of $T \sim 1.2\times10^4$ K and $T \sim 3.4\times10^4$ K, respectively. A second and consistent constraint on the electron temperature comes from \citet{brown2005}, who report temperatures for a set of H$\alpha$-detected bow shocks in the range of $6\times10^3$ K to $1.4\times10^5$ K. We will use the latter range as a reference in this analysis.

For all bow shocks considered in this Section, the E-BOSS catalogue lists inferred ISM densities based on the stellar wind properties and the measured standoff distance. We expect a density enhancement in the shock, by a factor 4 predicted by the Rankine-Hugoniot equations \citep{landau1959} up to a factor $\sim 10$ seen in bow shock simulations by \citet{gvaramadze2018}. In order to compare different bow shocks, we can therefore plot their electron density-temperature relation as measured from the radio image, $n_e(T; S_\nu)$, divided by their geometrically inferred surrounding ISM density, $n_{\rm ISM}$. We expect this over-density factor $n_e(T; S_\nu)/n_{\rm ISM}$ to lie, roughly, in the range between $1$ (i.e. no over-density) to $10$. 

\begin{figure}
 \includegraphics[width=\columnwidth]{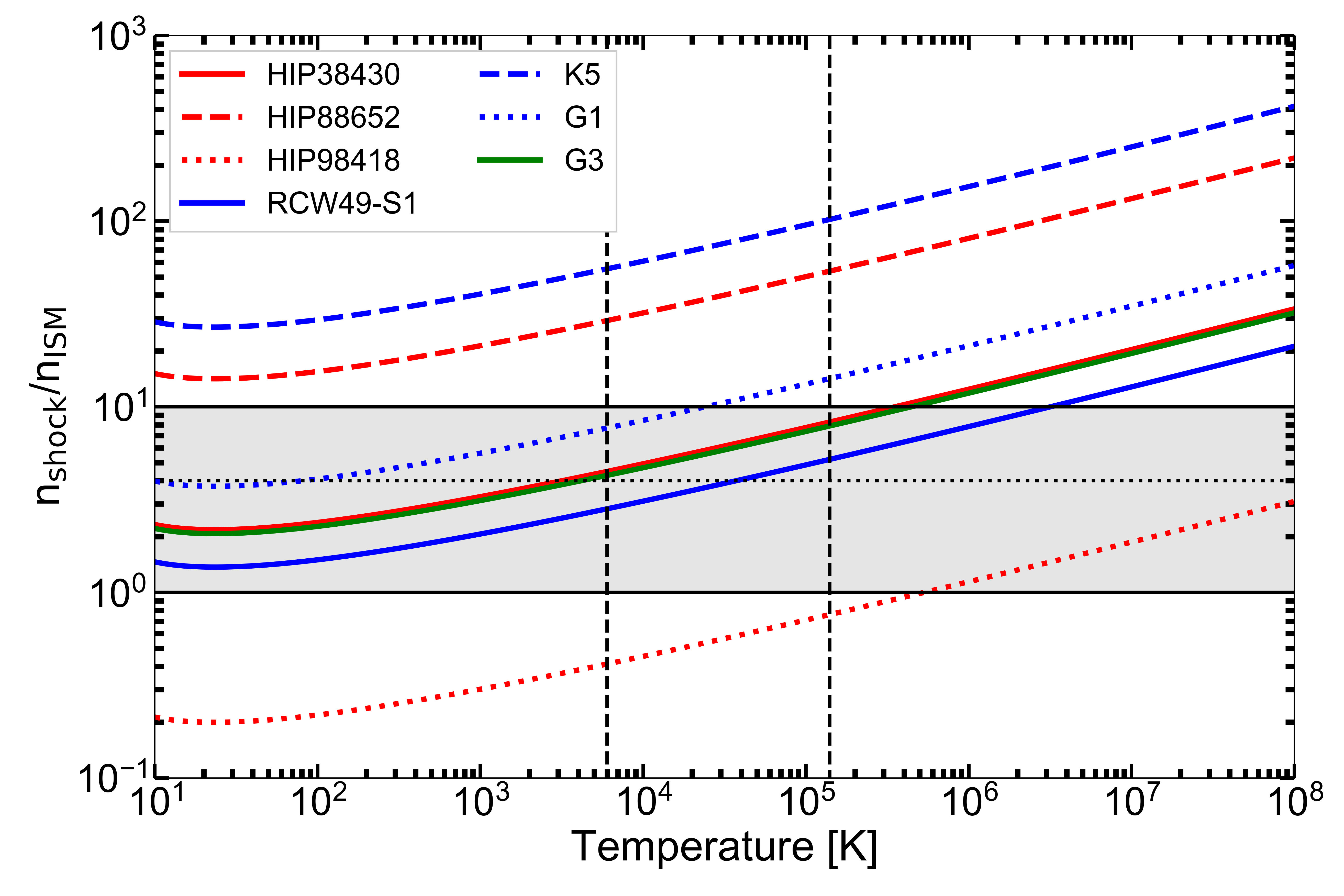}
 \caption{The relation between electron temperature and over-density to account for the peak radio flux density of the seven (candidate) radio bow shocks with known geometrical properties (i.e. excluding S2 and S3 in RCW 49). The grey region shows over-densities compared to the ISM between $1$ and $10$, while the dotted line indicates the theoretically expected value of $4$. The dashed black lines enclose the region in electron temperatures derived for a sample of H$\alpha$ bow shock by \citet{brown2005}. Four targets pass through the realistic region in this parameter space, enclosed by the black lines and dashed lines. Two sources (K5 and HIP 88652) appear to be too radio-bright, while HIP 98418 appears radio faint.}
 \label{fig:thermal}
\end{figure}

In the left panel of Figure \ref{fig:thermal}, we plot the electron over-density factor as a function of temperature for the radio-detected targets. The grey area indicates the expected over-density range, while the black dashed lines enclose the typical shock temperature range from \citet{brown2005}. Evidently, four out of the seven plotted sources pass through the region of the plot where the expected over-density and temperature overlap: HIP 38430, RCW 49 S1, G1, and G3. In other words, their detected peak radio emission can be explained via a thermal scenario. HIP 98418, shown with the red dotted line, lies at lower over-density factors; therefore, in order to be consistent with an overdense shock compared to the ISM, it should show higher temperatures than typically inferred from H$\alpha$ bow shock detections. Finally, HIP 88652 and K5 show over-densities in the range of $\sim 30$--$100$, exceeding the expectations for shocks discussed above. In other words, their peak radio flux density greatly exceeds the flux densities that may be expected for a thermal scenario.

Finally, for S1 in RCW 49, we can include the radio emission reported by \citet{benaglia2013} in $5.5$ GHz ATCA observations. \citet{benaglia2013} report a total flux density, integrated over the bow shock, of $70\pm10$ mJy. In the RACS images, we measure a peak brightness of $12.6 \pm 0.23$ mJy/beam, while the entire bow shock is approximately five times larger than the ASKAP beam. Therefore, we can roughly estimate a total RACS bow shock flux between $\sim 50$--$100$ mJy; differences between the precise definition of the bow shock size, radio telescope and array configurations, as well as the level of brightness fluctuations across the RACS radio bow shock, mean that a more precise estimate is not attainable at this stage. An integrated RACS flux density of that order of magnitude is consistent with a $\alpha = 0.1$ radio spectrum, expected for thermal emission. A non-thermal scenario ($\alpha \sim 0.5$) would imply values closer to $\sim 175$ mJy instead. We stress, however, that due to the difficulties in comparing different radio arrays at different frequencies, these estimates should be considered with caution.

\subsection{Non-detected bow shocks}

Finally, for the radio-non-detected bow shocks, we can use the non-thermal and thermal frameworks to estimate their expected peak radio flux densities in RACS observations. For this purpose, in the thermal scenario, we assume that the source either has a relatively weak magnetic field of $10$ $\mu$G or its maximum magnetic field $B_{\rm max}$. To obtain an optimistic prediction, we assume a high injection efficiency of $10$\%, in combination with $E_{\rm max}=10^{12}$ eV and $\alpha = 0.5$. For the thermal scenario, we assume either $T=6\times10^3$ K or $T=1.4\times10^5$ K and an over-density of $4$ -- as these flux density predictions scale with $n_e^2$, they can be multiplied by $2.5^2$ to consider over-densities of $10$.

\begin{figure}
 \includegraphics[width=\columnwidth]{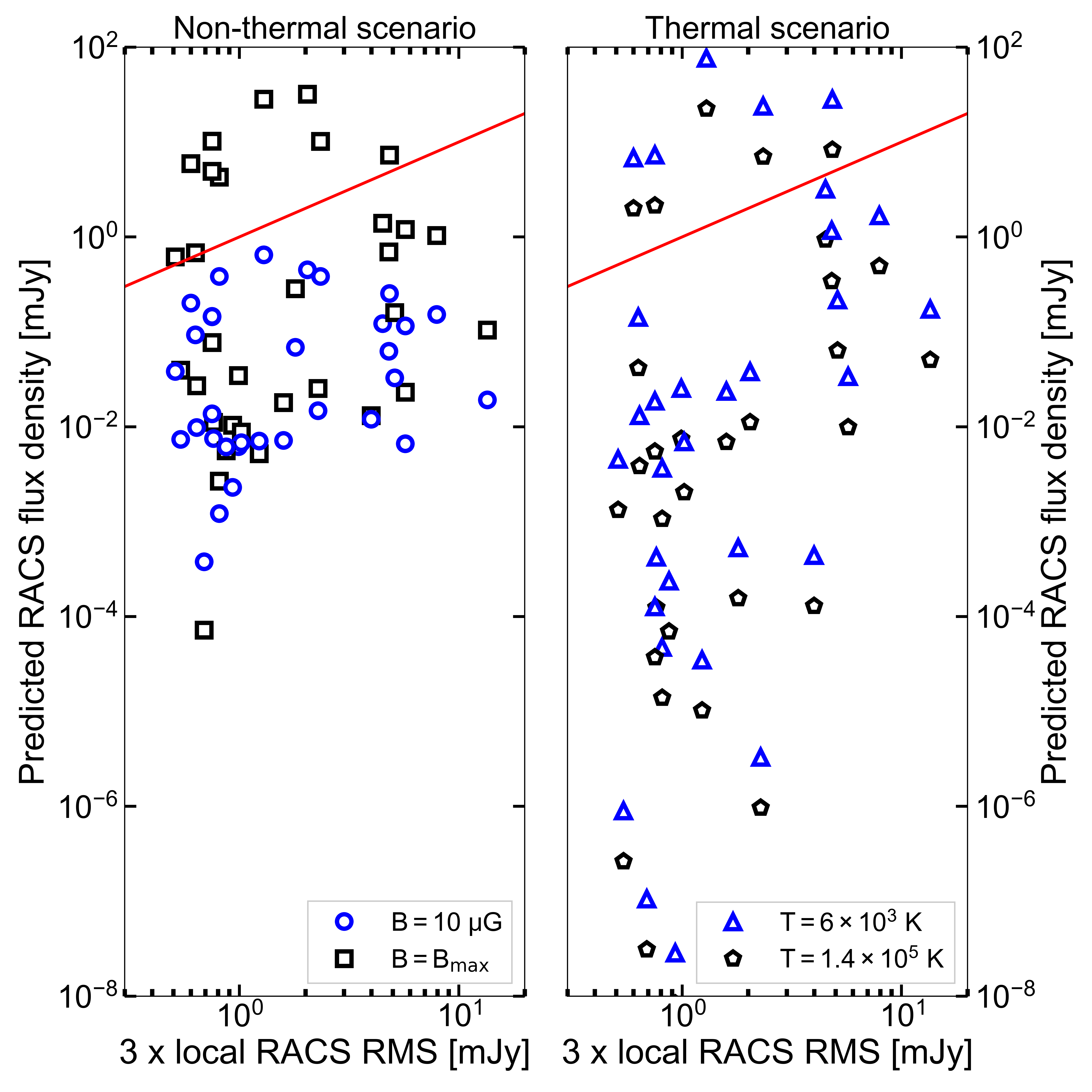}
 \caption{The predicted peak radio flux density in the RACS beam for radio-non-detected targets, plotted versus three times their local RMS sensitivity. The red line indicates the one-to-one line; sources above this line should, at least, show detectable emission in one beam. The left panel shows the predictions for two non-thermal scenarios, where a weakly-magnetized scenario ($B=10$ $\mu$G) leaves all sources undetectable. The right panel shows two thermal scenarios; in both, five targets would be expected to be detected. We discuss these sources in Section \ref{sec:future_exp}, where the non-thermal and thermal scenarios are also fully defined.}
 \label{fig:nondet}
\end{figure}

In Figure \ref{fig:nondet}, we plot the predicted RACS flux densities in the brightest beam for the non-thermal (left) and thermal (right) scenarios. The different markers indicate different magnetic fields (left) or temperatures (right). In both panels, the red line shows where the two flux densities are equal; points above this line may be expected to be detected with RACS. In the non-thermal scenario, all non-detections can be accounted for if not all bow shocks have their maximum magnetic field. As mentioned earlier, it would require exceptional fine-tuning for each source to be close to $B=B_{\rm max}$; therefore, this requirement is likely to be met. In the thermal scenario, however, it appears more complicated to explain all non-detections: for five sources (HIP 75095, HIP 77391, HD 57682, K7, and G4) thermal emission is expected to be detectable for both considered temperatures. A lower density enhancement or overestimated ISM density may move these sources below the sensitivity level -- none of the sources exceeds three times the local RMS for an electron density equal to the E-BOSS ISM density. We will, however, return explicitly to these five sources in the Discussion.

\section{Discussion}
\label{sec:5}

\subsection{The radio detections and their origin}

In this work, we report the RACS detection of three confident and three likely radio counterparts of runaway IR stellar bow shocks, as well as three inconclusive/to-be-confirmed candidates, out of a sample of 50 sources (e.g. Table \ref{tab:assessment} and Section \ref{ref:search}). We then performed simple analytical calculations to assess whether a self-consistent non-thermal or thermal emission scenario can account for the observed radio emission, in order to shed light on the emission's origin (Table \ref{tab:scenarios} and Section \ref{sec:calcs}). Finally, we briefly assessed how the IR bow shocks without associated radio emission fit into those thermal and non-thermal frameworks. 

The three confident radio detections of IR bow shocks -- G1 and G3 in NGC 6357, and S1 in RCW 49 -- increase the number of radio runaway stellar bow shocks from two \citep{benaglia2010,vandeneijnden2021} to five. All three of these fit well in a thermal scenario, where a free-free origin is responsible for the radio emission. In contrast, synchrotron emission appears to be unable to explain their detected radio luminosities, requiring unfeasibly high efficiencies of injecting energy into their relativistic electron population. All three sources are located in complex environments with large regions of diffuse radio emission and, possibly, large local variations in ISM densities. Therefore, one may conclude that indeed a complex ISM around the massive runaway star may play an important role in creating the circumstances where thermal radio emission becomes detectable. This conclusion also fits with the thermal origin for the radio bow shock of Vela X-1 proposed in \citet{vandeneijnden2021}, as this system may have recently crossed a local ISM over-density \citep{gvaramadze2018}. However, one should be careful: in the NGC 6357 field, only two out of the seven bow shocks that are visible to ASKAP are detected; in Cyg OB2, this ratio is one out of six. Complex ISM environments may therefore be beneficial or even necessary for thermal bow shock radio emission, but are clearly not sufficient; especially if these complex environments are highly structured and even closely-separated runaway stars encounter different structures.

The role of a complex ISM environment can also hamper or complicate the detection of radio bow shocks. In K5, the first of the three likely bow shock detections, neither a thermal nor a non-thermal scenario can explain the observed radio luminosity. Possibly, the bow shock emission is observed in superposition with emission from the large-scale diffuse structure to its South (Figure \ref{fig:HIP88652_K5_field_radvswise}, bottom left). As this diffuse structure shows similar flux densities to the shock region, it could explain why we observe a shock luminosity inconsistent with the energetics of the stellar wind and the density in the ISM. More extreme cases of this issue are present for S2 and S3 in RCW 49 -- two of the three aforementioned \textit{inconclusive} counterparts. While both show associated radio emission, their location inside a large-scale, diffuse radio structure prevents a clear identification of a bow-shock-like morphology. This issue may persist even at higher resolution or sensitivity, although spectral index maps may help to separate bow shock emission from other diffuse structures -- especially if any bow shock emission is, at least partially, non-thermal. 

That leaves three sources to discuss: HIP 98418, HIP 38430, and HIP 88652. The former two are likely detections, as argued in Section \ref{ref:search}. If real, HIP 98418 is best explained as a non-thermal source with a magnetic field above $\sim 30$ $\mu$G to ensure a reasonable injection efficiency (i.e. below $\sim 10$\%); it may include a thermal contribution, but it would require a high electron temperature to explain all emission through a free-free origin. For HIP 38430, if confirmed, both scenarios can explain the emission, individually or combined. Again, to ensure realistic injection efficiencies, the magnetic field should be relatively high at $B \gtrsim 60$ $\mu$G. This minimum value, naturally, decreases if thermal processes provide a significant portion of the radio luminosity. As we will return to in the discussion of non-detected sources, these results confirm how high magnetic fields (i.e. a significant fraction of the maximum) are essential to ensure non-thermal radio detections at current sensitivities. 

In the case of HIP 38430, it is also interesting to note how this source contains a stellar system of high multiplicity, likely consisting of at least two stellar binaries \citep{lorenzo2017}. This exotic nature may affect several of the assumptions underlying our calculations and bow shock morphology in general \citep{wilkin1996}: the cumulative wind outflow may not be spherical, the relevant wind kinetic power may be incorrectly estimated due to the presence of several massive stars, and the magnetic field strength may be affected by mergers in the system. Therefore, the exact derived quantities, such as magnetic field or injection efficiency, for this target should be treated with caution. We note that none of the other sources (possibly) detected in this paper have been reported to host binary or higher-multiplicity systems. 

In addition, the multiplicity of HIP 38430 introduces the possibility that a fraction of the radio emission may originate from colliding stellar winds, similar to colliding wind binaries (CWBs). The position of HIP 38430 is close to the edge of the bow shock region drawn in Figure  \ref{fig:HIP98418_38430field_radvswise}. For the E-BOSS distance to HIP 38430, the luminosity of the radio counterpart is consistent with other CWBs \citep{debecker2013}. Also, one of the binaries making up HIP38430 was recently found to be a wide, detached binary consisting of two O stars with likely masses of $30$ $\rm M_{\odot}$ each \citep{lorenzo2017}. Therefore, it is possible that colliding winds contribute to the observed radio counterpart. The system's multiplicity may also have a second effect: the E-BOSS distance of $900$ parsec, estimated from Hipparcos parallaxes, is lower than other estimates in the literature. For instance, \citet{lorenzo2017} argue for a distance closer to $\sim 5$ kpc, which would render both the non-thermal and thermal bow shock scenario less reasonable by increasing $\eta_e$ (scaling $\propto D^2$) and $n_e$ (scaling $\propto D$). Instead, at such larger distances, the radio luminosity of HIP 38430 is similar to the peak luminosity of the radio-brightest CWB Apep at similar wavelengths \citep{callingham2019}. This scenario can be tested directly through radio follow-up monitoring of HIP 38430: the radio luminosity of CWBs is highly variable along their eccentric binary orbits due to variations in the distance between the massive stars.

Finally, for HIP 88652, the non-thermal scenario can explain the emission for a magnetic field $B \gtrsim 50$ $\mu$G. The bow shock identification of this source is, however, less confident than that of the aforementioned two sources. From the RACS image, we can see how image artefacts due to two close-by, bright point-like sources strongly affect the region where the bow shock is expected to lie. Therefore, one can wonder whether the brightest beam is representative of the entire region here. While the peak flux density is $2.2$ mJy/beam, the mean value is $\sim 0.9$ mJy/beam over the region drawn in Figure \ref{fig:HIP88652_K5_field_radvswise} (top; we remind the reader that these regions are mainly intended to guide the eye). Given that the imaging artefacts increase the mean flux density in the region, we may expect that any emission for the bow shock could be less than $\sim 0.9$ mJy/beam. In that scenario, injection efficiencies in the non-thermal scenario below $10$\% could be possible at lower magnetic fields. Therefore, to assess this effect, a dedicated observation of this field with ASKAP or MeerKAT, or alternatively a re-imaging of the existing data to remove the image artefacts, is needed to confirm the presence and assess the nature of this radio bow shock. 

Using NVSS images at $1.4$ GHz, \citet{peri2012} and \citet{peri2015} also searched for radio emission associated with the bow shocks in the two E-BOSS catalogues. Radio emission was reported at the positions of G2, G3, SER 5, S1, S3, HIP 88652, and HIP 38430 (as well as HIP 11891, which was not covered by RACS). However, these radio sources did not always fully overlap with the bow shock shape and position, nor could the authors conclude whether the radio emission was associated with the bow shock. Out of these radio candidates, we confirm the presence of radio emission at the positions of G3, S1, HIP 88652, and HIP 38430. In all cases, the reported NVSS flux densities are significantly (5 to 8 times) higher than the RACS peak flux densities. Given the small difference in observing frequency, such differences are unlikely to be dominated by spectral shape, but instead likely caused by the lower spatial resolution of NVSS, different array configuration, and possible positional offsets; the NVSS beam is likely to cover the entirety or a significant fraction of the bow shock, thereby observing its integrated radio flux density. In addition, given the complexity of the direct surroundings of G2, S3, HIP 88652, and HIP 38430, the NVSS images of these shocks may have also picked up (significant) contributions from surrounding diffuse radio emission regions. This comparison highlights the importance of the enhanced spatial resolution of RACS in order to determine the morphology of a possible radio bow shock and to separate it from other radio structures; however, the RACS non-detection of G2, which showed possible faint radio emission at lower resolution in NVSS \citep[at $4\pm1$ mJy/beam;][]{peri2015}, also reminds one how such enhanced resolution is only useful if accompanied by sufficient sensitivity to prevent emission being resolved out.  

\subsection{H$\alpha$ emission in the thermal scenario?}
\label{sec:halpha}

For the four targets where the thermal emission can realistically account for the (likely) observed radio bow shocks (G1, G3, S1, and HIP 38430), we can consider whether H$\alpha$ line emission from the same thermal electron population may be expected. We can apply the formalism introduced by \citet{gvaramadze2018} and applied in \citet{vandeneijnden2021} to estimate the H$\alpha$ surface brightness for a given temperature, electron density, and bow shock depth. Following our assumption that the width and depth of the bow shock are similar, and using the theoretical over-density factor of four, we find that all four targets would show a surface brightness between $\sim 60$ and $\sim 600$ Rayleigh\footnote{1 Rayleigh $\equiv 5.66\times10^{-18}$ erg s$^{-1}$ cm$^{-2}$ arcsec$^{-2}$.} ($T\approx10^5$ K) or $\sim 500$ and $\sim 5000$ Rayleigh ($T\approx10^4$ K). The Vela X-1 bow shock, for comparison, was detected by \citet{gvaramadze2018} at a surface brightness of $43$ Rayleigh in the SuperCOSMOS H$\alpha$ Survey \citep{parker2005}\footnote{\url{http://www-wfau.roe.ac.uk/sss/halpha/index.html}}. Therefore, one automatically wonders whether these four bow shocks show up in the SuperCOSMOS images as well.  

\begin{figure*}
 \includegraphics[width=\textwidth]{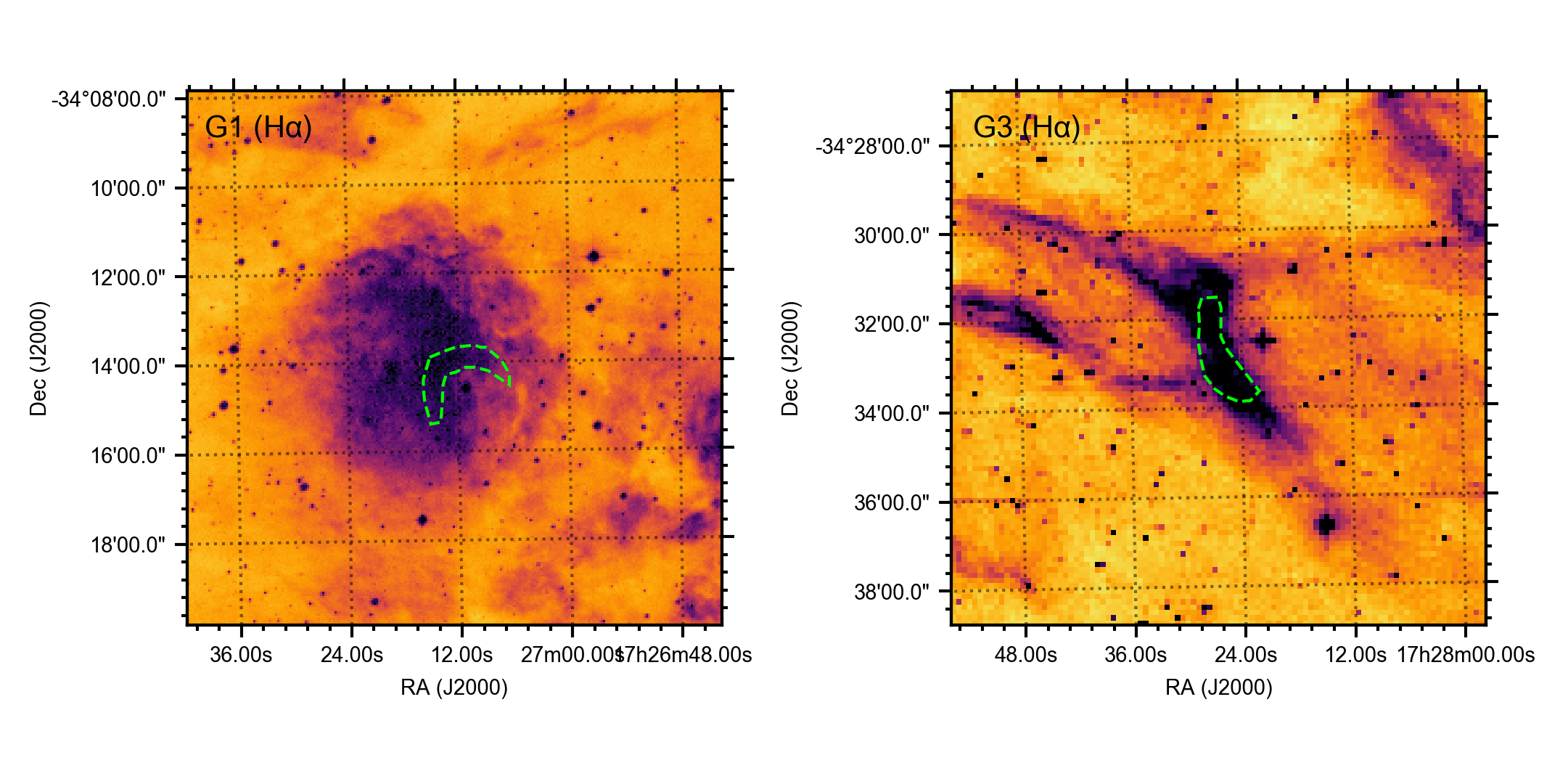}
 \caption{The SuperCOSMOS H$\alpha$ Survey images of G1 (left) and G3 (right), with the same contours and on the same physical scale as their radio and IR images in Figure \ref{fig:Gfield_radvswise}. H$\alpha$ emission is clearly visible for both fields, but only for G3, it clearly traces out a bow-shock like morphology, providing supportive evidence for a thermal scenario underpinning its radio emission. In G1, larger-scale H$\alpha$ emission complicates the search for a bow shock, although hints of a cavity may be present to the South/South-West of the drawn region.}
 \label{fig:halpha}
\end{figure*}

In Figure \ref{fig:halpha}, we show the SuperCOSMOS H$\alpha$ images for G1 and G3, overlaid with the same contours as in their radio and IR images (Figure \ref{fig:Gfield_radvswise}). In the case of G3 (right), a clear diffuse H$\alpha$ structure can be seen tracing the position (and extending slightly further) of the IR and radio bow shock. This H$\alpha$ detection is strong supporting evidence of a thermal emission origin for the radio emission. In G1 (left), a diffuse region is observed with significantly larger extent than the bow shock. However, hints of a cavity may be present to the South/South-West of the indicated bow shock region, toward the position of the star. In HIP 38430, similar issues arise: the large-scale diffuse structure also seen in radio (and even weakly in IR; Figure \ref{fig:HIP98418_38430field_radvswise}) dominates the image. However, since it is saturated completely, we cannot assess whether there is any emission associated with the likely radio bow shock. These two cases highlight the issue with searching for bow shocks in H$\alpha$ discussed previously by \citet{brown2005} and \citet{meyer2016}, namely the presence of larger-scale HII regions that can surround the runaway star. Finally, S1 does not show a clear, diffuse H$\alpha$ structure in the shape of a bow shock; however, the field is complex, with both many point sources and other diffusive structures, possibly complicating its detection. 

\subsection{Future expectations: thermal versus non-thermal detectability}
\label{sec:future_exp}

In this work, the majority of radio and candidate radio bow shocks appear to fit better in a thermal rather than non-thermal scenario. The latter scenario has been considered extensively in the literature, due to its possible ability to yield non-thermal sources of very-high energy emission via inverse Compton scattering of IR or stellar photons \citep{delvalle2012,schulz2014,toala2016,toala2017,debecker2017}. This possibility has motivated searches for radio emission as signatures of such non-thermal electron populations \citep[e.g.,][]{peri2015,benaglia2021}, but the results in this work introduce the question whether the radio band is the best place to search: does the free-free/thermal emission complicate this approach?

To assess this question, we turn again to the radio-non-detected sources from the E-BOSS catalogues. For those where the ISM density and geometrical properties are known, we can estimate and compare their free-free and synchrotron emission. Such a comparison will always include a large number of assumptions, for instance regarding the over-density factor and electron temperature (for free-free emission) or injection efficiency, spectral index, and magnetic field (for synchrotron emission). Therefore, we cannot make one unique comparison, but instead consider two extreme scenarios. In the \textit{free-free-favourable case}, we assume a density of four times the ISM and a low temperature ($T=6\times10^3$ K), combined with a low magnetic field of $10$ $\mu$G and an injection efficiency of $1$\%. Alternatively, in the \textit{synchrotron-favourable case}, we instead assume that $T=1.4\times10^5$ K, while we change the magnetic field to its maximum value for each source and increase $\eta_e$ to $10$\%. All other parameters remain the same; in both cases, we assume $p=2$ ($\alpha=0.5$) and $E_{\rm max} = 10^{12}$ eV.

\begin{figure}
 \includegraphics[width=\columnwidth]{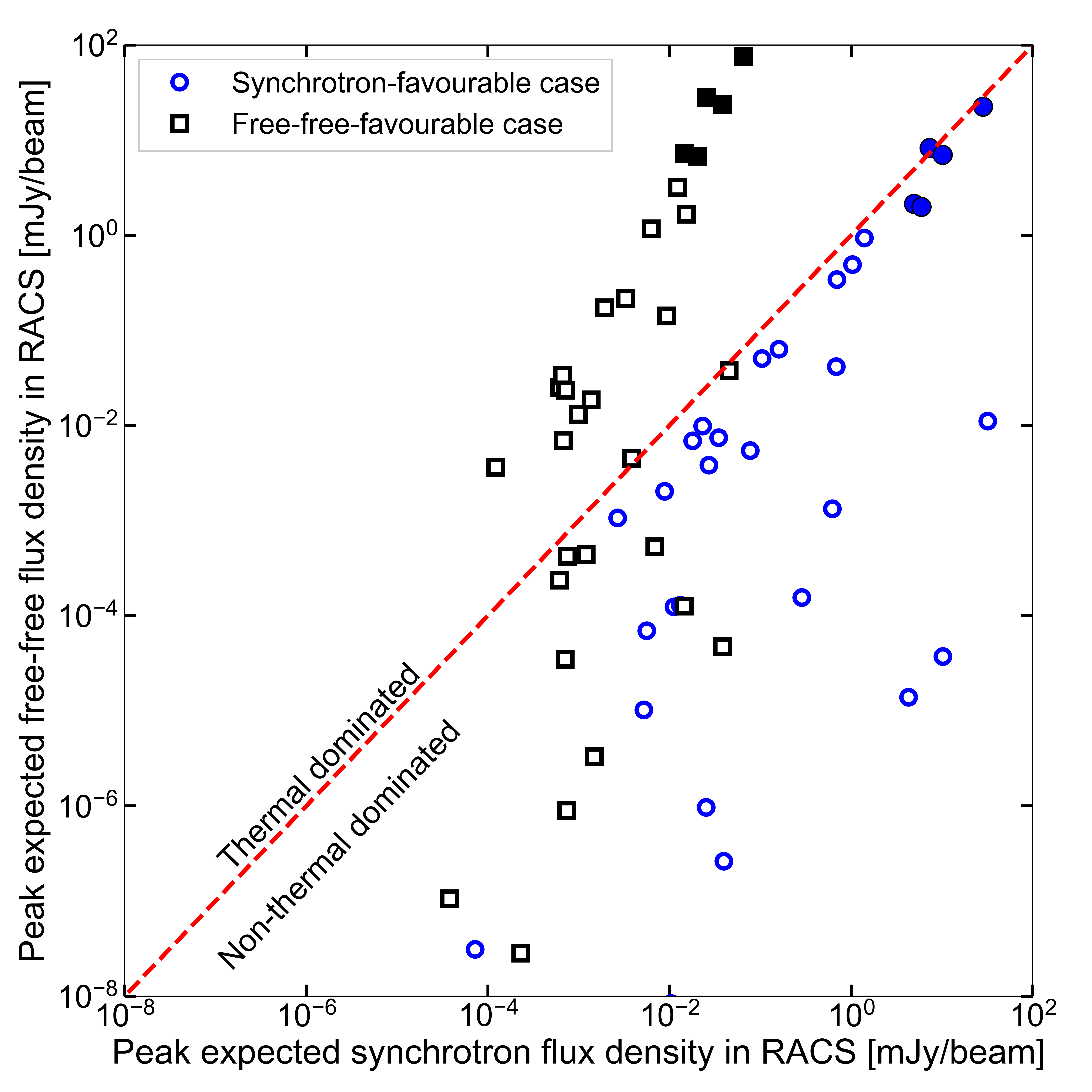}
 \caption{Comparison between the expected RACS free-free and synchrotron flux density, for free-free-favourable (black squares) and synchrotron-favourable (blue circles) circumstances (see the text for details). In the former scenario, the majority of sources is dominated by thermal emission with a weaker synchrotron contribution. In the latter case, however, the majority of sources lies close to equal contributions of synchrotron and free-free emission (the red dashed line). The filled points show the five sources that showed detectable thermal radio emission in the right panel of Figure \ref{fig:nondet}.}
 \label{fig:comparison}
\end{figure}

In Figure \ref{fig:comparison}, we plot the predicted thermal and non-thermal RACS radio flux densities against each other; the position of a source with respect to the red dotted line shows which of the two processes dominates. In the synchrotron-favourable case, the synchrotron emission is indeed the main source of emission, although in many cases the free-free emission is barely fainter -- despite quite extreme parameters for this scenario, especially regarding the magnetic field. In the free-free-favourable case, however, the thermal emission significantly dominates over non-thermal emission in the majority of sources. Those sources where this is inverted, are typically faint and less likely detectable. Only a few sources reach flux densities above $1$ mJy/beam in RACS, where they may become detectable with recognizable morphologies. Those are predominantly the filled points, corresponding to the five sources in the right panel of Figure \ref{fig:nondet} -- i.e. the five non-detected targets where realistic thermal emission may have been expected to be detected in RACS.

This leaves us with two conclusions and a note: firstly, dominant or significant thermal/free-free emission may be expected in most radio-detected bow shocks, given the stellar and ISM properties of the sources in E-BOSS; only extreme non-thermal properties ($B$, $\eta_e$, or available stellar wind kinetic power) appear to result in dominant synchrotron emission. This argues for the determination of reliable radio spectral index maps of these bow shocks as an additional determinant between the two scenarios, especially if one aims to find evidence for non-thermal emission. Secondly, for the five outlier sources in the thermal scenario, our assumed properties likely do not hold; the ISM density or stellar wind properties may be incorrect, or the depth may be significantly less than the bow shock width. Finally, we note that while we have treated the two scenarios independently, their detectability is not: for instance, a higher ISM density increases the shock density and thermal emissivity, but also decreases the standoff distance and thereby increases the stellar wind kinetic power passing through the shock region. This may lead, at higher sensitivity, to the detection of shocks displaying a combination of both types of emission. 

\subsection{Future recommendations}

We briefly turn to the assumptions made in the non-thermal calculation and the role for new observations in testing those. In the bottom panel of Figure \ref{fig:nonthermal}, we show the effect of varying the spectral index (or, more explicitly, $p$) and the maximum electron energy. This shows how especially variations in $p$ can lead to significantly higher required injection efficiencies. The maximum energy, on the hand, does not have a large quantitative effect. More implicitly, we assume in our analysis that, apart from synchrotron emission, other non-thermal cooling processes \citep[i.e. inverse Compton scattering, relativistic Brehmsstrahlung;][]{delvalle2012,delpalacio2018} have negligible effect. Similarly, we assume that, even though synchrotron emission is the dominant non-thermal emission process, it does not dominate the electron losses. Instead, we assume that such losses are dominated by escape, setting the maximum electron energy and injection efficiency. 

Most analyses in the literature, for BD+43$^{\rm o}$3654 \citep{benaglia2010,benaglia2021} and Vela X-1 \citep{vandeneijnden2021} as well as other studies focused on high-energy bow shock emission \citep[e.g.][]{delvalle2012,schulz2014,debecker2017}, show how synchrotron loss time scales approach diffusive escape time scales only at the highest electron energies for typical bow shock properties. However, with more sensitive follow up observations, such calculations can be performed in detail to test these assumption for the sources in this work. In addition, if spectral indices can be measured, the injection efficiency and therefore the non-thermal scenario can be assessed more confidently. Finally, such observations can also help to understand whether the bow shock has reached steady state -- an underlying assumption in all our calculations \citep[see][]{mohamed2012} -- by measuring the shape of the shock more accurately than IR observations and comparing that to the steady-state shape model by \citet{wilkin1996}.

Two other obvious recommendations for future studies are two expansions of this work: expansion of the number of sources and of the number of Stokes parameters. In this search, we have focused on the E-BOSS catalogue, as it lists ISM and stellar wind properties, as well as distances, for most targets. However, if one is only to perform a search, without necessarily detailed follow up calculations, larger catalogues such as that by \citet{kobulnicky2016} can be consulted. By including polarization studies, currently only performed for BD+43$^{\rm o}$3654 \citep{benaglia2021}, we obtain an additional approach to understand the emission mechanism -- beyond emissivity and spectral shape. Such studies likely require pointed observations to reach good sensitivity to low levels of polarization, and targeted data analysis, as publicly available surveys tend to present the Stokes I data. 

We end this Discussion by emphasizing how, for the two first radio-detected and best-studied runaway stellar bow shocks BD+43$^{\rm o}$3654 and Vela X-1, sensitive, high-resolution, pointed radio observations have been essential to perform detailed calculations of their energetics, emission processes, and polarization properties. BD+43$^{\rm o}$3654 showed hints of arc-shaped structure in NVSS images, but dedicated VLA observations where required to confirm its presence \citep{benaglia2010}. The much fainter radio bow shock of Vela X-1 does not show up in RACS (RMS of $222$ $\mu$Jy), but was serendipitously detected in pointed MeerKAT observations (RMS of $40$ $\mu$Jy). The advent of new observatories, such as MeerKAT, ASKAP, and the future SKA and ngVLA, will likely lead to the detection of more radio bow shocks or strong candidates in survey data, similar to this work. With dedicated follow-up observations of such sources and candidates, we expect that a large sample of well-studied radio bow shocks can be gathered, to ultimately shed light on the particle acceleration efficiencies and high-energy emission of such shocks, as well as their thermal emission properties.

\section{Conclusions}
\label{sec:conc}

In the work, we have searched for radio emission associated with the massive runaway stellar bow shocks in the E-BOSS catalogue, using the Rapid ASKAP Continuum Survey. In these radio images, we find bow shock counterparts for three bow shocks (G1 and G3 in NGC 6357, and S1 in RCW 49), three strong candidates (HIP 98418, HIP 38430, and K5 in Cyg OB2), and three candidates requiring follow-up observations (S2 and S3 in RCW 49, and HIP 88652). For the seven bow shocks where sufficient geometrical and ISM properties are known to perform simple emission calculations, we find that four (G1, G3, S1, HIP 38430) are likely dominated by free-free emission -- although in the latter, synchrotron emission may also contribute significantly. Alternatively, the bow shocks of HIP 98418 and HIP 88652 appear to be dominated by synchrotron emission, while, finally, K5 remains difficult to explain in either scenario. There, instead, we may observe a significant contamination from other diffuse structures, while image artefacts complicate the identification and calculations in HIP 88652.

These radio detections and candidates significantly increase the number of radio counterparts to massive runaway stellar bow shocks, which was only two before this work. Therefore, we also assess whether, with future arrays and surveys, this sample is expected to grow further. Given the expected flux densities, it likely is, although whether those sources will be dominated by thermal or non-thermal processes is challenging to predict. Finally, we recommend that the new counterparts and candidates presented in this paper are studied in more detail with pointed observations, to fully map out their structure, measure their radio spectral index maps, and their polarization. With such detailed information, the nature of the candidate counterparts can be tested, and the nature of the emission can be investigated in more detail. Ultimately, such studies will therefore help to understand to what extent and with what efficiencies, these bow shock can act as particle accelerators and sources of non-thermal very-high energy emission. 




\section*{Acknowledgements}


We thank the referee for a constructive report that improved the quality and clarity of this work. This paper includes archived data obtained through the CSIRO ASKAP Science Data Archive, CASDA (\url{https://data.csiro.au}). The Australian SKA Pathfinder is part of the Australia Telescope National Facility (grid.421683.a) which is managed by CSIRO. Operation of ASKAP is funded by the Australian Government with support from the National Collaborative Research Infrastructure Strategy. ASKAP uses the resources of the Pawsey Supercomputing Centre. Establishment of ASKAP, the Murchison Radio-astronomy Observatory and the Pawsey Supercomputing Centre are initiatives of the Australian Government, with support from the Government of Western Australia and the Science and Industry Endowment Fund. We acknowledge the Wajarri Yamatji people as the traditional owners of the Observatory site. This work makes use of several \textsc{python} packages, namely \textsc{numpy} \citep{oliphant_numpy}, \textsc{astropy} \citep{astropy13,astropy18}, \textsc{matplotlib} \citep{hunter07}, and \textsc{aplpy} \citep{robitaille12}. JvdE is supported by a Lee Hysan Junior Research Fellowship awarded by St. Hilda's College, Oxford. 

\section*{Data Availability Statement}

A Jupyter notebook to generate the figures in this work and redo the analysis in the Discussion, can be found on this link upon publication: \url{https://github.com/jvandeneijnden/RACSRadioBowshocks}. The RACS images are available at \url{https://research.csiro.au/racs/home/data/}. WISE data can be accessed via \url{https://irsa.ipac.caltech.edu/applications/wise/}, while SuperCOSMOS data is available at \url{http://www-wfau.roe.ac.uk/sss/halpha/}.





\input{paper.bbl}

\bsp	
\label{lastpage}
\end{document}